\documentclass[10pt]{article}
\usepackage{fix-cm}
\usepackage{evox_report}
\usepackage{graphicx}
\usepackage{booktabs}
\usepackage{float}
\usepackage{needspace}
\usepackage[section]{placeins}
\usepackage[hidelinks]{hyperref}
\usepackage{url}
\usepackage[
  backend=biber,
  style=authoryear-comp,
  dashed=false,
  sorting=nyt,
  giveninits=true,
  uniquename=init,
  maxbibnames=20,
  maxcitenames=2,
  mincitenames=1,
  doi=true,
  url=true,
  isbn=false
]{biblatex}

\usepackage{longtable}
\usepackage{microtype}

\DeclareNameAlias{sortname}{family-given}

\let\cite\parencite
\newcommand{\gnum}[1]{\textbf{#1}}

\title{Persistent recursive worlds enable autonomous software evolution}
\author{%
\textbf{Beichen Huang} \quad \textbf{Zhenyu Liang} \quad \textbf{Bowen Zheng} \quad \textbf{Ran Cheng}$^{*}$\\[0.30em]
Department of Data Science and Artificial Intelligence, The Hong Kong Polytechnic University\\
Hong Kong SAR, China\\[0.18em]
$^{*}$Correspondence: \href{mailto:ranchengcn@gmail.com}{ranchengcn@gmail.com}
}
\hypersetup{pdftitle={Persistent Recursive Worlds Enable Autonomous Software Evolution},pdfauthor={Beichen Huang, Zhenyu Liang, Bowen Zheng, Ran Cheng}}
\AtBeginBibliography{\urlstyle{same}}
\begin{document}
\maketitle

\begin{genesisabstract}
Software systems accumulate dependencies and design commitments over time, while the agents that modify them operate within bounded sessions. How, then, can development remain cumulative across generations of transient agents? We introduce \texttt{EvoX Genesis}, which organizes development as a \textbf{persistent recursive software world}. Agents enter an accepted repository version, delegate work within local scopes, and submit changes for verification and parent review. Once accepted, those changes pass to later episodes together with the contracts and tests used to assess them. The project thus preserves both earlier work and a basis for evaluating what comes next. We examine this continuity through four studies of formation, continuation and redevelopment. Starting without compiler code, Genesis built a C compiler with \textbf{219,676 lines of Rust source} through \textbf{1,019 agent episodes} over 123.4 hours. Within this growing implementation, repairs increased passes on a fixed set of 100 Csmith programs from 61 to 93; a subsequent revision \textbf{retained all 93 passes}. Development also continued in a separate compiler lineage when DeepSeek V4 Flash inherited a GLM~5.2 snapshot. To examine continuity when the implementation itself changes, Genesis redeveloped selected MESA functionality in Rust, with numerical agreement assessed against the Fortran reference across six workloads. The WebGL trajectory, meanwhile, exposed both repairs and regressions as repeated checks recorded progress towards \textbf{2,017/2,019} conformance pages passed. A complementary EvoX Compiler case connects program-merging performance to executable graph and backend contracts. Together, these records trace how accepted work becomes a foundation for further development: later agents inherit both usable implementations and concrete ways to assess further revisions.
\end{genesisabstract}

\section{Introduction}
Software tasks end; software systems endure. As a project grows, new contributions must work within module boundaries, interface contracts and architectural commitments established by earlier development~\cite{parnas1972modules,lehman1980programs}. They must also preserve behaviour captured by existing regression suites, so that added functionality and repairs survive subsequent changes. \emph{Development is cumulative when earlier contributions remain usable as a foundation for what comes next.}

Large language models (LLMs) now allow coding agents to perform much of this work autonomously. Agents inspect repositories, execute tools, propose modifications and run tests, while benchmarks increasingly cover sequential upgrades and repository synthesis alongside isolated issue resolution~\cite{jimenez2024swebench,yang2024sweagent,wang2024openhands,thai2025sweevo,xu2026roadmapbench,ding2025nl2repo}. As development extends across successive tasks, however, the project's dependencies and history grow beyond the context and execution time of any individual agent. Later contributors must work with the consequences of earlier decisions, including interface drift or defects left by locally successful patches~\cite{orlanski2026slopcodebench,yan2026slump}. Preserving the project therefore requires a way to carry its implementation and behavioural expectations from one episode to the next. This mismatch between a persistent project and its transient contributors motivates our central question: \emph{how can software development remain cumulative across generations of transient agents?}

We address this question with \texttt{EvoX Genesis} (hereafter, \texttt{Genesis}), an autonomous software evolution system organized as a \emph{persistent recursive world}. Each episode begins from an accepted repository version and a local directory path, which together establish the state it inherits and the scope of its work. Within that version, agents can delegate recursively; returned changes enter project history after verification and parent review. The resulting implementation, directory contracts and executable checks then pass to subsequent episodes, giving new agents both material to extend and criteria against which to assess their changes. In this way, the project maintains continuity as its contributors come and go. This separation between transient contributors and persistent project records resonates with Popper's account of objective knowledge: theories, arguments and problem situations remain available for others to examine and extend~\cite{popper1972objective}.

We examine this continuity through four studies that vary what agents inherit and what they must preserve. Compiler formation first traces how local contributions assemble into an interdependent system and how repairs survive later revisions. Once a working implementation exists, the question becomes whether fresh agents can continue from it, including after a change of foundation model; a separate compiler lineage provides this continuation study. MESA redevelopment then examines what can be carried forward when the implementation language itself changes, using the inherited numerical reference to assess the replacement. Software WebGL formation follows repairs and regressions through successive conformance and rendering checks, while the complementary EvoX Compiler case examines transformations whose performance can change even when executable interface requirements remain fixed. Across these settings, repository histories connect accepted contributions to subsequent work, and repeated evaluations reveal which tested behaviours persist.

The report makes three contributions:
\begin{itemize}
  \item We formalize the persistent recursive world using version and path coordinates, separating delegation within an accepted state from integration that advances project history.
  \item We implement this organization in \texttt{EvoX Genesis}, with bounded manager and executor episodes, isolated Git worktrees, and versioned directory contracts that provide inherited context.
  \item We document four development studies and an EvoX Compiler case, connecting accepted contributions to subsequent work through repository ancestry, repeated functional checks, recorded repairs and executable interface contracts.
\end{itemize}

\Needspace{12\baselineskip}
\section{Related work}
\subsection{Coding agents and extended software development}
Coding agents operating at repository scale couple LLMs with file inspection, command execution, code editing and automated test feedback. SWE-bench, SWE-agent and OpenHands established issue-level evaluation and general agent platforms~\cite{jimenez2024swebench,yang2024sweagent,wang2024openhands}. Evaluation has since expanded to sequential version upgrades and feature evolution~\cite{thai2025sweevo,xu2026roadmapbench,shastry2026swesteps}, as well as repository synthesis from natural language specifications~\cite{ding2025nl2repo}. These longer trajectories bring dependencies between edits into focus: shifting interfaces, diverging assumptions and accumulating technical debt can degrade agent reliability as work proceeds~\cite{orlanski2026slopcodebench,yan2026slump}. \texttt{Genesis} addresses this continuity at the project level by maintaining an accepted software world across more than a thousand bounded episodes.

\subsection{Memory, coordination and persistent project state}
Agentic systems preserve continuity through several forms of memory and coordination. Reflection loops and adaptive memories retain selected past trajectories, skill libraries archive reusable procedures, and multiagent frameworks organize work through role hierarchies and conversational workflows~\cite{shinn2023reflexion,gao2026swemem,wang2024voyager,hong2024metagpt}. Project notes outside the active prompt and explicit execution ledgers extend this continuity beyond the immediate conversation~\cite{yan2026slump,gloaguen2026agentsmd,wang2026ledger}. In \texttt{Genesis}, inherited context takes the form of versioned directory contracts within a shared repository. Agents use those contracts during recursive delegation, while parent review determines which returned changes become part of the state inherited by later episodes.

The repository also supplies a durable record of accepted changes and their ancestry. EvoGit uses this version structure to let decentralized agents mutate and recombine code across a Git graph without centralized message passing or shared conversational state~\cite{huang2025evogit}. Within a version, however, contributors still need boundaries that make local work compatible with the larger project. Modular information hiding bounds change propagation~\cite{parnas1972modules}, and nearly decomposable hierarchies contain local perturbations~\cite{simon1962architecture}; both are relevant to software that must continue changing over time~\cite{lehman1980programs}. \texttt{Genesis} brings these spatial and temporal concerns together by instantiating agents recursively across directory paths within an accepted version. Shared contracts support coordination through persistent project records~\cite{theraulaz1999stigmergy}, while tests and parent review determine which returned changes advance that version.

\subsection{Program evolution and scientific software redevelopment}
Iterative generation and selection provide a related route to algorithmic discovery. FunSearch evolves program kernels against explicit scalar evaluators, AlphaEvolve extends evaluator-guided search to broader algorithmic structures, and ERA optimizes scientific scripts against quantitative scores~\cite{romeraparedes2024funsearch,novikov2025alphaevolve,aygun2026era}. Building on the EvoX ecosystem~\cite{huang2025evox}, \texttt{Genesis} applies iterative development to one continuing software architecture, where accepted changes must coexist with the components already in place.

Scientific redevelopment makes this requirement concrete: a replacement implementation must preserve the numerical relationships on which scientific use depends. Such preservation is difficult outside the software's original environment~\cite{trisovic2022code,moreau2023containers}, making provenance and numerical verification central to redevelopment~\cite{barker2022fair4rs,kanewala2014testing}. The risks of unverified AI-generated changes further motivate checks that remain available through later revisions~\cite{obrien2025threats}. We therefore use Modules for Experiments in Stellar Astrophysics (MESA)~\cite{paxton2011mesa} to examine numerical behaviour during language migration.

\Needspace{17\baselineskip}
\section{EvoX Genesis: persistent recursive worlds}
\subsection{World coordinates and project state}
\texttt{Genesis} organizes autonomous software engineering around an accepted project that persists across agent episodes. At each developmental step, an agent enters a local software world defined by two orthogonal coordinates:
\begin{equation}
w=(v,p),
\label{eq:localworld}
\end{equation}
where $v$ denotes an accepted software version and $p$ a directory path within the repository. \emph{Version fixes temporal heritage; path bounds spatial responsibility.} Agents at distinct paths within $v$ share its complete accepted state and historical lineage. An agent at $(v,p)$ can read across the repository to resolve dependencies, while its primary context assembly and write authorization are anchored at $p$. The accepted state comprises source code, directory contracts (\texttt{CONTEXT.md}), test suites, reusable skills and provenance records.

\begin{figure}[!tbp]
\centering
\includegraphics[width=0.99\linewidth]{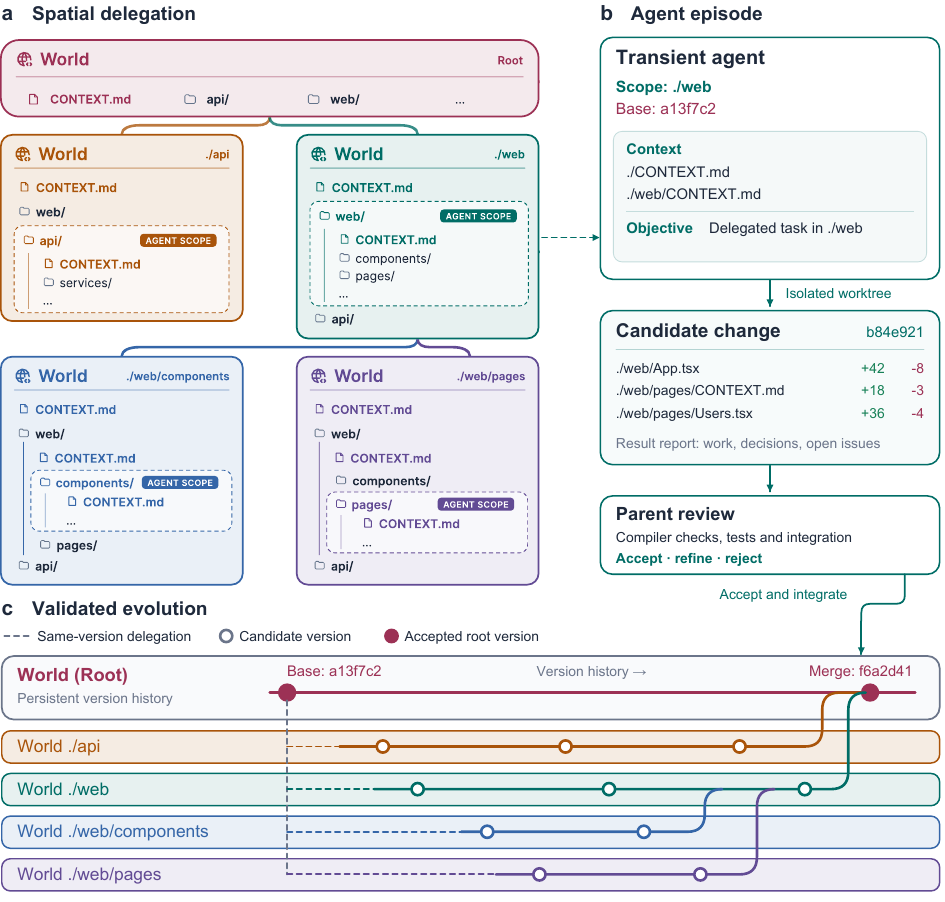}
\caption{\textbf{Persistent recursive worlds decouple spatial delegation from temporal evolution.}
\textbf{a}, An accepted software version $v$ can be entered from distinct paths within the repository, defining local worlds $w=(v,p)$. The version $v$ fixes the accepted project state and evolutionary lineage, whereas the path $p$ bounds where an agent begins and what scope it governs. Spatial recursive delegation $(v,p)\rightsquigarrow(v,q)$ instantiates a child agent at subpath $q$ while keeping the accepted version $v$ invariant.
\textbf{b}, A transient agent receives context and a bounded local objective $g_i$, executes in an isolated worktree, and returns a candidate change $\Delta_i$ with a result report. The responsible parent accepts, requests refinement or rejects the candidate using compiler checks, tests, architectural constraints and integration evidence.
\textbf{c}, Dashed connectors link delegated scopes to the same accepted base version $v$; open nodes mark candidate versions and filled nodes mark accepted root versions. Accepted changes propagate through parent worlds and advance persistent version history via $(v,p)\rightarrow(v',p')$. Commit hashes and file changes are illustrative.}
\label{fig:architecture}
\end{figure}

\subsection{Transient agents and recursive delegation}
Given an episode objective $g_i$ at local world $(v,p)$, a transient agent $A_i$ executes a bounded sequence of model and tool turns within an isolated Git worktree to propose a candidate modification:
\begin{equation}
\Delta_i=A_i\bigl((v,p),g_i\bigr).
\label{eq:transientagency}
\end{equation}
During this episode, $A_i$ maintains private conversational and scratchpad state. That state expires at termination, so subsequent agents resume from an accepted world coordinate.

Work that spans several modules can be divided through spatial recursive delegation. A parent agent at $(v,p)$ instantiates a child at subpath $q$, shifting the local scope while preserving the accepted version:
\begin{equation}
(v,p)\rightsquigarrow(v,q).
\label{eq:delegation}
\end{equation}
The child can delegate further within scope $q$, with $v$ held fixed throughout. Leaf \emph{executor} agents synthesize and test code; root and intermediate \emph{manager} agents partition objectives, maintain \texttt{CONTEXT.md} interface contracts and audit returned evidence. Each bounded episode thus handles work at its own hierarchical level, while changes to the shared project await acceptance.

\Needspace{14\baselineskip}
\subsection{Validation and persistent lineage}
A candidate modification $\Delta_i$ enters the persistent software world through an accepted software event,
\begin{equation}
(v,p)\longrightarrow(v',p'),
\label{eq:acceptedevent}
\end{equation}
which advances the global version history from $v$ to $v'$ (where $p'=p$ unless an accepted structural move, rename or deletion maps the path to $p'$). The responsible parent reviews the returned patch and result report against compiler diagnostics, regression suites, scope constraints and integration checks. Acceptance makes the patch part of the version inherited by subsequent episodes; rejection leaves $v$ unchanged. Findings from a rejected attempt can still inform later work when they are committed to path contracts, regression tests or provenance records.

The Elixir/OTP runtime implements this inheritance by assembling prompts from versioned \texttt{CONTEXT.md} files along the path from the repository root to the active directory $p$. Git commits and protected archive references preserve the accepted state, while isolated Git branches and ZFS clone worktrees hold candidate changes during execution. When an episode ends, its BEAM supervision state and temporary worktree can be discarded; subsequent episodes recover their context from the accepted project.

\section{Evaluation}
\subsection{Evaluation protocol and accounting}
We examine cumulative development by following what happens to earlier work as new episodes begin. Repository ancestry identifies contributions retained in the accepted project; continuation from an inherited snapshot shows how later agents use that project; repeated functional checks assess behaviour across revisions. The four studies cover \emph{de novo formation}, \emph{continuation across model replacement} and \emph{scientific redevelopment}, using the same persistent recursive world architecture with distinct initial conditions and verification oracles. The EvoX Compiler case extends this examination to program-merging performance and the graph and backend contracts retained at v0.1.0 (Section~\ref{sec:evoxc-main}).

Before execution, operators define the root task objectives, external validation suites, tool sandboxes and controller ceilings (maximum recursion depth, turn limits and context compression thresholds). They specify milestones and initiate root phase transitions as development proceeds. In the compiler formation study, the specification defined target C11 semantics, CLI compatibility, target instruction sets (x86 and x86-64) and external test suites without supplying seed compiler code or directory layouts; development spanned two sequential root phases connected by an automated repository handoff. In the continuation study, both lineages received the identical prebuilt GLM~5.2 compiler world and a single continuation instruction. In the MESA redevelopment study, operators provided the immutable Fortran reference repository, the target scope of 13 modules and the numerical benchmark harness. For these three studies, runtime archives and repository histories record agent objectives, lineage trees, accepted commits and token accounting; prelaunch setup and root phase transitions are documented at the protocol level. The WebGL study uses a versioned verification report, implementation contracts and repair commits.

Throughout our analysis, an \emph{archived episode} denotes one completed transient task record in the runtime archive; total \emph{spawned agent} counts exceed archived episode counts when interrupted or unarchived worker processes occur. A contribution is classified as \emph{retained} when its Git commit resides in the ancestor graph of the final accepted repository HEAD; retention records integration ancestry, while the validation suites measure tested functionality. For the archived compiler and MESA runs, elapsed wall time measures real time from root launch to terminal commit; cumulative agent hours sum overlapping episodes. The WebGL development interval is the span between the first and last agent commits. Monetary figures reflect direct foundation model API token charges; the WebGL value is reported by the project record. Physical line counts (\texttt{LoC}) are measured deterministically across tracked text files under experiment manifests and include comments and blank lines. Complete protocol, archive and measurement specifications are provided in the Supplementary Information.

\FloatBarrier
\subsection{Formation: a C compiler from scratch}
\paragraph{Setup.}
The first experiment follows cumulative development from a repository with no implementation to inherit. The specification requested \emph{jcc}, an independent C compiler in Rust for LLVM workflows, requiring a command-line interface compatible with Clang, standard object-file and linker integration, LLVM IR export, C11 as the primary target, and mandatory x86 and x86-64 back ends. Direct translation or linking of Clang or LLVM source code was prohibited. The run used DeepSeek V4 Flash with \texttt{xhigh} reasoning effort and a context compression threshold of 150,000 tokens. Phase~1 started from commit \texttt{41e087ce90f3}, containing only \texttt{.gitignore} and \texttt{genesis.toml}; Phase~2 inherited the generated repository at commit \texttt{04d64b89d504} together with a phase handoff summary. External verification suites comprised c-testsuite, LLVM test programs, LZ4 and SQLite, alongside Csmith differential test programs generated and executed by the committed harness.

\paragraph{Results.}
During formation, root and intermediate managers partitioned the specification across repository paths, transient executors synthesized local modules in isolated worktrees, and parent nodes gated integration against automated test suites. Over \gnum{123.4 hours} of elapsed operation, the run archived \gnum{1,019 agent episodes} (out of 1,065 spawned agents) and reached delegation depth five (Supplementary Fig.~\ref{fig:supp-compiler-lineage} and Tables~\ref{tab:compiler-lineage-accounting} to \ref{tab:compiler-phases}). The final repository (commit \texttt{a8c116ec7ed3}) contained \gnum{248,989 physical lines} across 750 tracked text files (including 219,676 lines of Rust across 354 \texttt{.rs} files; Supplementary Fig.~\ref{fig:supp-compiler-composition} and Table~\ref{tab:compiler-file-types}), at a recorded model token cost of \gnum{US\$44.38}.

\begin{figure}[!htbp]
\centering
\includegraphics[width=\linewidth]{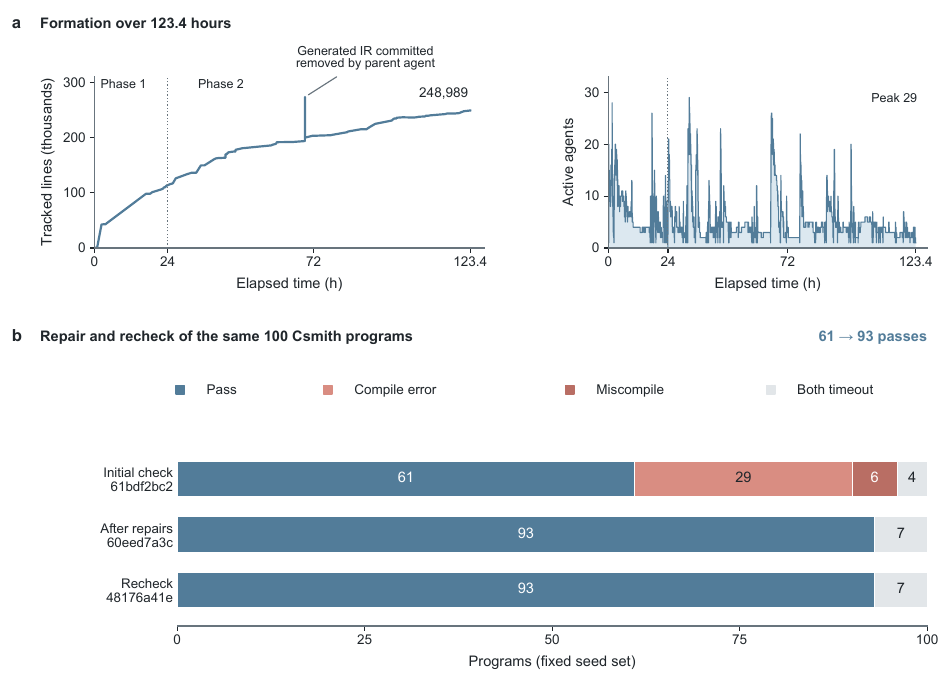}
\caption{\textbf{Compiler formation accumulates code and retains verified repairs.}
\textbf{a}, Tracked text lines and concurrent agents over 123.4 hours. Dotted lines mark the phase handoff at 23.905 hours. Line counts include all Git-tracked text files, comments, blank lines and generated text artifacts. At 69.115 hours, an agent inadvertently committed a 74,107-line compiler-generated intermediate representation (IR) dump. Its parent agent removed the generated file 3 seconds later, producing the transient spike.
\textbf{b}, Three evaluations of the same 100 Csmith programs at the labelled compiler commits. Repairs retain the 61 initial passes and convert six miscompiles and 26 compilation errors into passes. The other three compilation errors become timeouts under both jcc and Clang. The recheck retains all 93 passes and the same seven timeouts (Table~\ref{tab:evidence-sources}, \protect\hyperlink{evidence:jcccsmith2026}{E01}). Both later evaluations reuse the original programs, with 600-second compilation and 300-second execution limits.}
\label{fig:compiler-development}
\end{figure}

The final compiler passed \gnum{220/220} reported c-testsuite cases and \gnum{32/36} evaluated LLVM cases; the four remaining LLVM cases corresponded to documented GNU extension policy gates. Differential testing yielded \gnum{93} Csmith passes from 100 seeds, with seven programs timing out under both compilers. Integration and workspace checks added 8/8 LZ4 checks, 2/2 SQLite shell compilation and deterministic query stages, \gnum{2,904} Rust workspace unit tests and 106/106 internal regression cases (Fig.~\ref{fig:compiler-development} and Table~\ref{tab:compiler-validation}).

The committed Csmith records allow these final outcomes to be followed back through the repair process (Table~\ref{tab:evidence-sources}, \protect\hyperlink{evidence:jcccsmith2026}{E01}). Initially, 61 of the same 100 programs passed, 29 produced compilation errors, six were miscompiled and four timed out under both jcc and Clang. Repairs subsequently converted all six miscompiles and 26 compilation errors into passes. The remaining three programs also compiled successfully, but timed out under both compilers. A subsequent check retained all 93 passes and the same seven timeout cases, showing that the repaired behaviour survived a later revision (Fig.~\ref{fig:compiler-development}b; Table~\ref{tab:csmith-repair-stages}).

\Needspace{8\baselineskip}
\paragraph{Interpretation.}
Compiler development depends on earlier representation choices: syntax and type information constrain intermediate code, optimization and register allocation, while linking brings separately developed components together. As these dependencies accumulate, Git ancestry records how local contributions enter the shared implementation. Among the 1,015 records in the reconstructed lineage, 929 (91.53\%) contributed commits retained in the final ancestry; four archived records with missing parents are excluded from this denominator (Table~\ref{tab:compiler-lineage-accounting}). The fixed-program rechecks add behavioural evidence to that history by showing that repaired functionality survived a subsequent revision. Taken together, the two records trace how earlier episodes supplied a working foundation for later implementation and repair.

\FloatBarrier
\subsection{Continuation: a compiler across foundation models}
\paragraph{Setup.}
Having examined how a compiler forms, we next ask whether its accumulated implementation and tests support further development after the original agents terminate, including when the foundation model changes. Both continuation branches started from a lineage separate from the DeepSeek formation run: the completed \emph{jcc} repository built by GLM~5.2 at commit \texttt{37216cfa254a} (504 archived initial episodes; Supplementary Fig.~\ref{fig:supp-continuation-lineages} and Table~\ref{tab:continuation-settings}). They received the same starting repository, committed context contracts, continuation prompt, verification suites and controller ceilings on the same dedicated host. One branch continued with GLM~5.2 (to commit \texttt{ab58a91460ad}), whereas the other replaced the foundation model with DeepSeek V4 Flash (to commit \texttt{10077f656861}).

\begin{figure}[!htbp]
\centering
\includegraphics[width=\linewidth]{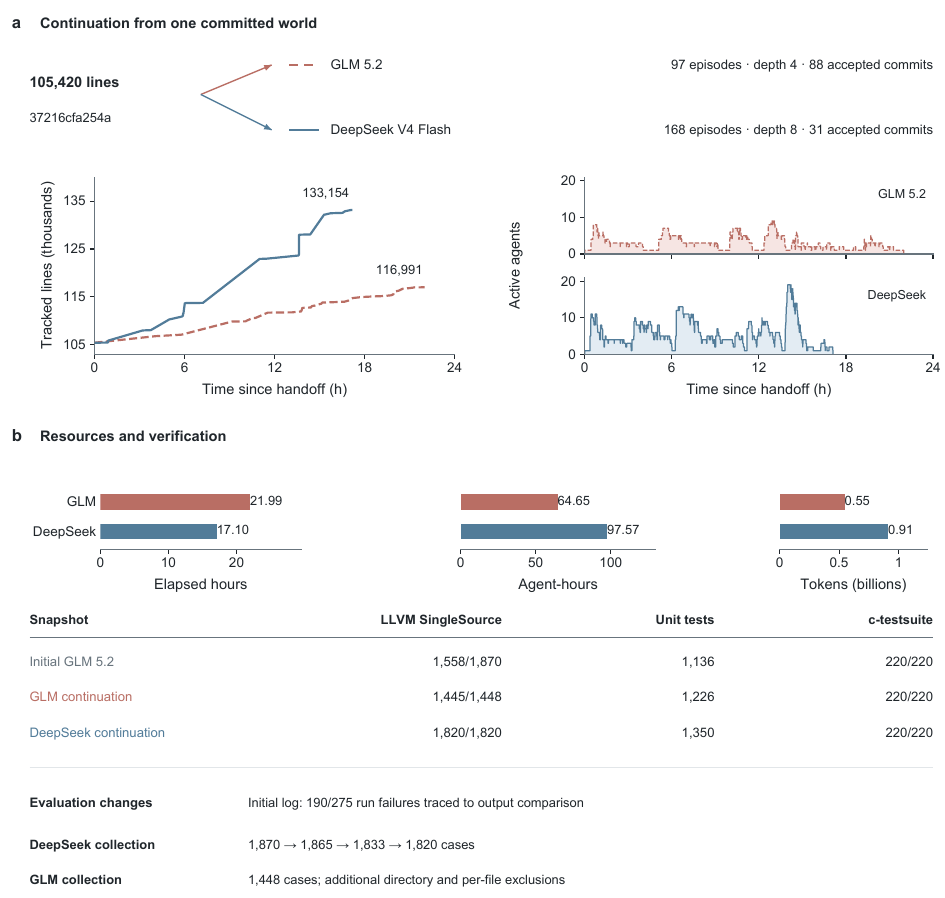}
\caption{\textbf{One committed compiler supports two continuation trajectories.}
\textbf{a}, Both branches inherit the same 105,420-line repository. Growth and concurrent activity use common time scales. Dashed salmon denotes GLM~5.2; solid blue denotes DeepSeek V4 Flash. Counts give archived episodes, maximum delegation depth and accepted first-parent commits.
\textbf{b}, Resource use and final verification at each snapshot. LLVM fractions use each snapshot's retained case set; workspace tests and c-testsuite retain separate units. The lower rows summarize the recorded output-comparison defect and subsequent collection changes (Table~\ref{tab:evidence-sources}, \protect\hyperlink{evidence:jccllvmaudit2026}{E02}, \protect\hyperlink{evidence:jccdsprotocol2026}{E03}, \protect\hyperlink{evidence:jccglmprotocol2026}{E04}). Line counts include tracked Rust, Cargo.toml, Markdown and shell files.}
\label{fig:model-switchable-continuation}
\end{figure}

\paragraph{Results.}
Both lineages resumed development directly from the committed GLM~5.2 repository and extended its implementation and test infrastructure through newly instantiated transient agents, with no conversational handoff from the original sessions (Fig.~\ref{fig:model-switchable-continuation}, Supplementary Figs.~\ref{fig:supp-continuation-organization} to \ref{fig:supp-continuation-lines} and Tables~\ref{tab:continuation-summary} to \ref{tab:continuation-lines}). GLM~5.2 continuation spawned 98 agents (97 archived records), reached delegation depth four ($d=4$), expanded workspace unit tests from 1,136 to 1,226 and passed 1,445/1,448 cases in its retained LLVM SingleSource suite. Continuation with DeepSeek V4 Flash spawned 178 agents (168 archived records), reached the controller ceiling of delegation depth eight ($d=8$), expanded workspace unit tests to 1,350 and passed \gnum{1,820/1,820 (100\%)} cases in its retained LLVM SingleSource manifest. Although the branches began from the same snapshot, their retained test manifests evolved during continuation. Their final pass fractions therefore refer to their respective case sets, while token consumption and elapsed time also differed under the shared controller ceilings.

Changes to those manifests accompanied repairs to the evaluation machinery itself. A committed audit of the initial 1,870-case log traced 190 of 275 run failures to handling of the reference output's trailing exit marker (Table~\ref{tab:evidence-sources}, \protect\hyperlink{evidence:jccllvmaudit2026}{E02}). DeepSeek's collected set subsequently changed from 1,870 to 1,865, 1,833 and 1,820 cases through documented exclusions covering test helpers, target architecture, reference outputs and unsupported features. GLM retained 1,448 cases under its own directory and file filters (Table~\ref{tab:evidence-sources}, \protect\hyperlink{evidence:jccdsprotocol2026}{E03}, \protect\hyperlink{evidence:jccglmprotocol2026}{E04}). The final results thus reflect both continued compiler development and the evaluation protocol retained by each branch; Supplementary Section~\ref{sec:continuation-evaluation-changes} details the protocol changes.

\paragraph{Interpretation.}
Both model families resumed work from the inherited implementation and test infrastructure through fresh agent episodes. Their common starting state subsequently developed into distinct trajectories, with different delegation depths ($d=4$ versus $d=8$), concurrency profiles, commit granularity and code churn. Continuity here consists in the ability to take up and extend the same committed project after agent turnover and model replacement, with functional outcomes assessed against each branch's final retained suite.

\FloatBarrier
\Needspace{8\baselineskip}
\subsection{Redevelopment: MESA from Fortran to Rust}
\paragraph{Setup.}
The third experiment shifts from extending an inherited implementation to replacing its language while preserving numerical behaviour. Modules for Experiments in Stellar Astrophysics (MESA)\footnote{MESA source repository: \url{https://github.com/MESAHub/mesa/}} is a widely used open Fortran suite for stellar evolution~\cite{paxton2011mesa}. Treating a frozen MESA snapshot at commit \texttt{461dcba94f33} as an immutable reference world, we tasked \texttt{Genesis} (powered by DeepSeek V4 Flash) with reimplementing 13 core numerical and physics module directories into a cohesive Rust workspace (\texttt{mesa-rs}). The selected foundation and physics directories span \gnum{139,414 physical lines of Fortran}, including module tests; the 13-module scope excludes application engines such as \texttt{star}, \texttt{astero} and \texttt{binary}. The Rust \texttt{eos} and \texttt{kap} crates implement selected functionality from the corresponding MESA modules.

\Needspace{6\baselineskip}
Numerical behaviour and runtime were assessed across six standalone astrophysical workloads: nuclear burn proxy, equation of state (EOS) lookup, radiative opacity lookup, bicubic 2D interpolation, stiff Rosenbrock (ROS2) ODE integration and nonlinear root finding via Newton's method. Timing uses the median of 25 pinned executions after warmup per workload.

\begin{figure}[!htbp]
\centering
\includegraphics[width=\linewidth]{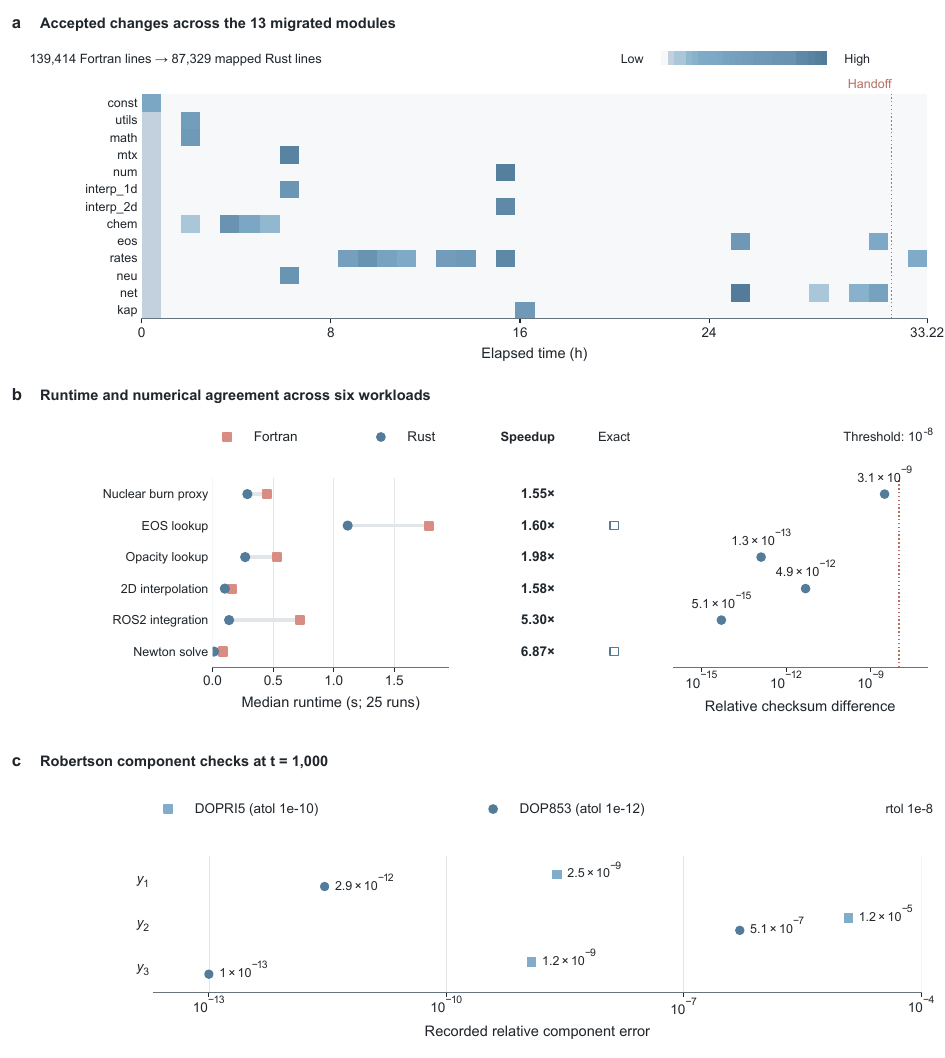}
\caption{\textbf{MESA redevelopment joins module integration, runtime and component checks.}
\textbf{a}, Accepted first-parent changes across 13 migrated crates over 33.22 hours. Time bins encode $\log(1+\mathrm{added}+\mathrm{deleted})$ lines; pale bins contain no accepted changes at this resolution. The dotted line marks the root handoff at 31.72 hours.
\textbf{b}, Median runtimes over 25 pinned executions, recorded Fortran/Rust ratios and checksum discrepancies. Open squares denote identical workload checksums; the threshold is $10^{-8}$. Ratios use unrounded timings.
\textbf{c}, Source-recorded Robertson component errors at $t=10^3$ for two Rust solvers, relative to the committed high-accuracy reference (Table~\ref{tab:evidence-sources}, \protect\hyperlink{evidence:mesarobertson2026}{E06}). Both use relative tolerance $10^{-8}$ and the labelled absolute tolerances. These component checks have their own reference and protocol (Supplementary Section~\ref{sec:mesa-component-checks}).}
\label{fig:mesa-redevelopment}
\end{figure}

\paragraph{Results.}
\texttt{Genesis} completed the scoped redevelopment in \gnum{33.22 hours} (commit \texttt{2ad071e658f8}), spawning 272 agents across 34 context paths (maximum depth $d=4$; Supplementary Figs.~\ref{fig:migration-module-lines} to \ref{fig:supp-migration-context} and Tables~\ref{tab:migration-settings} to \ref{tab:migration-modules}) at a recorded token cost of \gnum{US\$10.64}. The Rust workspace comprises \gnum{89,946 physical lines of Rust}: 67,374 library lines, 19,955 crate test lines and 2,617 root test and benchmark lines. It passed all \gnum{1,052} workspace unit tests with zero failures (Table~\ref{tab:migration-modules}).

To assess what this replacement preserved, we compared the Rust implementation with the Fortran reference on the six numerical workloads. EOS lookup and Newton workload checksums \gnum{match exactly}; relative checksum discrepancies across the other four range from $5.1\times10^{-15}$ to $3.1\times10^{-9}$. Under the same recorded build and host setup, these workloads also ran faster in Rust, with median speedups of \gnum{1.55$\times$ to 6.87$\times$} (Fig.~\ref{fig:mesa-redevelopment}b and Tables~\ref{tab:migration-performance} to \ref{tab:migration-dependencies}). A separate batch of 40 burn proxy runs yielded a median speedup of 1.23$\times$ (Table~\ref{tab:migration-deep-validation}).

\paragraph{Component and coupling checks.}
The ROS2 workload checksum measures mass conservation by summing the three species over 50 integrations. To assess the species individually, the committed Robertson tests compare their trajectories with a separate numerical reference (Table~\ref{tab:evidence-sources}, \protect\hyperlink{evidence:mesaworkloads2026}{E07}, \protect\hyperlink{evidence:mesarobertson2026}{E06}). At $t=10^3$, the recorded DOPRI5 component errors range from $1.2\times10^{-9}$ to $1.2\times10^{-5}$, and DOP853 errors from $1.0\times10^{-13}$ to $5.1\times10^{-7}$, using the absolute tolerances in Fig.~\ref{fig:mesa-redevelopment}c. The tests also track the transient second-species peak and compare against a high-accuracy reference with source-recorded SciPy cross-checks. An EOS-to-network integration test additionally checks conservation, energy accounting and provider call counts (Supplementary Section~\ref{sec:mesa-component-checks}).

\paragraph{Precision diagnosis.}
Pointwise interpolation tests provide a closer view of the arithmetic behind numerical differences (Table~\ref{tab:precision-diagnosis}) (Table~\ref{tab:evidence-sources}, \protect\hyperlink{evidence:mesagolden2026}{E05}). On a $31\times35$ grid, six evaluation points produce 18 bicubic values and first derivatives. The source diagnosis traces a maximum pointwise relative difference of $3.05\times10^{-8}$ against historical golden output to implicitly single-precision Hermite intermediates in Fortran, while Rust evaluates the path in double precision. When the Fortran transcription uses explicit double precision, its values match Rust exactly to the bit. The tests accordingly distinguish the historical golden comparison, which uses a $10^{-7}$ relative tolerance, from coordinate and coefficient checks that require exact equality. These pointwise golden checks complement the workload checksum measurements in Fig.~\ref{fig:mesa-redevelopment}b; their definitions and thresholds appear in Table~\ref{tab:precision-diagnosis}.

\paragraph{Interpretation.}
Here, continuity rests on the inherited numerical reference even as the implementation changes language and data structures. Workload checksums assess agreement in aggregate outputs, while component and pointwise comparisons examine the quantities and arithmetic paths beneath them. The precision diagnosis illustrates why both levels are useful: a discrepancy against historical output can be traced to the precision of intermediate calculations. Together, these checks document the audited behaviour carried into Rust and provide a basis for evaluating subsequent revisions.

\FloatBarrier
\subsection{Formation: a software WebGL renderer}
\paragraph{Setup.}
The fourth study examines another form of behavioural continuity: maintaining API conformance and rendered output as a software WebGL renderer develops in TypeScript. Its committed architecture separates shader compilation and linking, rasterization, WebGL state and objects, and presentation and image decoding. Directory contracts specify shared program representations, vertex and fragment interfaces, and error handling across these modules (Table~\ref{tab:evidence-sources}, \protect\hyperlink{evidence:webglarchitecture2026}{E10}). The project record reports DeepSeek V4 Flash and US\$36 in model token charges. The first and last agent commits span 65.9 hours, from 26 to 29 August 2026 (Table~\ref{tab:evidence-sources}, \protect\hyperlink{evidence:webglevidence2026}{E11}). Verification uses the committed JavaScript bundle \texttt{c4ffbff}, a browser harness and the two manifests in the recorded WebGL CTS 2.0.1 (beta) tree.

\paragraph{Results.}
The final verification report records \gnum{833/835} WebGL1 pages and \gnum{1,184/1,184} WebGL2 pages passed, giving \gnum{2,017/2,019 (99.90\%)} in total (Table~\ref{tab:evidence-sources}, \protect\hyperlink{evidence:webglfinal2026}{E09}). The software renderer was active on every graded page, and both suites completed with zero timeouts and zero runner errors. The two nonpassing WebGL1 pages concern low-alpha colour readback and double-underscore shader identifiers; the report records matching failures in native browser controls.

\begin{figure}[!htbp]
\centering
\includegraphics[width=\linewidth]{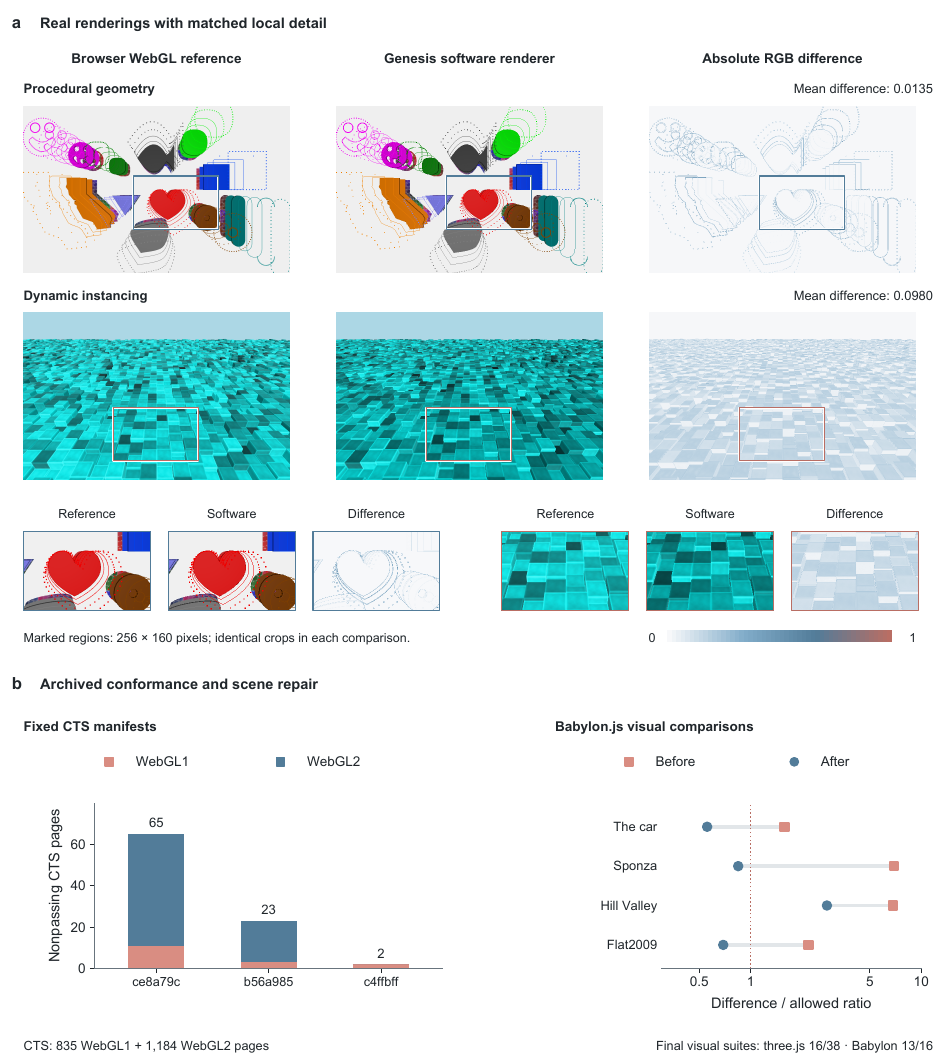}
\caption{\textbf{A software renderer connects scene output to conformance and repair.}
\textbf{a}, New captures of two three.js r184 examples using the committed \texttt{c4ffbff} renderer and Chromium WebGL with ANGLE SwiftShader. Each pair shares camera, geometry and instance inputs. Framed regions identify matched $256\times160$ pixel crops. Difference maps and means use $\Delta=\max_c|I_c^{\mathrm{software}}-I_c^{\mathrm{reference}}|/255$ for RGB channels on a common linear scale.
\textbf{b}, Archived CTS failures across three bundles, on fixed manifests of 835 WebGL1 and 1,184 WebGL2 pages. Paired Babylon.js measurements divide each scene's differing-pixel ratio by its acceptance budget; the dotted line marks that budget (Table~\ref{tab:evidence-sources}, \protect\hyperlink{evidence:webglstages2026}{E15}, \protect\hyperlink{evidence:webglbabylon2026}{E13}). The car uses 5\%; the other displayed scenes use 1.1\%. Capture and archived-suite protocols appear in Supplementary Sections~\ref{sec:webgl-visual-records} and \ref{sec:webgl-captures}.}
\label{fig:webgl-development}
\end{figure}

\Needspace{6\baselineskip}
This final result follows earlier complete runs on the same 2,019-page manifests, which passed 1,954 pages at bundle \texttt{ce8a79c} and 1,996 at \texttt{b56a985} (Table~\ref{tab:evidence-sources}, \protect\hyperlink{evidence:webglstages2026}{E15}). Figure~\ref{fig:webgl-development} places this trajectory alongside scene renderings and image-based repairs; Tables~\ref{tab:webgl-conformance} and~\ref{tab:webgl-repairs} give page totals and subtest counts for two resolved WebGL2 failures.

\paragraph{Rendered output.}
API conformance establishes how the renderer responds to specified calls; image comparisons then examine the output produced by complete scenes. The archived evaluations report 16/38 selected three.js pages and 13/16 selected Babylon.js scenes passing their respective image criteria (Table~\ref{tab:evidence-sources}, \protect\hyperlink{evidence:webglevidence2026}{E11}). To make the remaining visual differences inspectable, Fig.~\ref{fig:webgl-development}a provides a separate, fixed-version comparison of procedural geometry and dynamic instancing using the frozen software renderer. Matching scene inputs and common-scale pixel difference maps expose discrepancies in boundaries and shading; Supplementary Section~\ref{sec:webgl-captures} specifies the browser, scene version and capture protocol.

The archived Babylon comparisons also follow scene output through repair, linking changes in the renderer to changes in the images it produces. The differing-pixel ratio falls from 7.64\% to 0.93\% for Sponza, from 7.89\% to 2.78\% for The car, and from 2.40\% to 0.76\% for Flat2009. Hill Valley likewise improves from 7.54\% to 3.08\%, although it remains above its 1.1\% acceptance budget (Table~\ref{tab:evidence-sources}, \protect\hyperlink{evidence:webglbabylon2026}{E13}). The associated repairs address pre-blend clamping, matrix operand aliasing and the texture level-of-detail footprint.

\paragraph{Repair trajectory.}
The origin-clean failure arose from API validation order: texture binding was checked before the image source's security status. Commit \texttt{7d8718a} moved the origin-clean check ahead of binding validation at all six image-source entry points, changing the affected page from 9 passing and 4 failing subtests to 13 passing subtests. A second repair preserved 16-bit PNG input through image decoding and corrected RGBA channel handling, changing the high-bit-depth texture page from 2 passing and 1 failing subtests to 3 passing subtests. The subsequent full run raised the WebGL2 result from 1,182/1,184 to 1,184/1,184 while retaining the previous passes. This preservation followed an earlier period in which improvements and regressions occurred together. During the \texttt{ce8a79c} to \texttt{b56a985} transition, WebGL2 recovered 39 previously failing pages and introduced five new failures, giving a net gain of 34 passes (Table~\ref{tab:evidence-sources}, \protect\hyperlink{evidence:webglstages2026}{E15}). By retaining the same manifests across revisions, the evaluations make both recovery and regression visible, so subsequent repairs can be judged against the behaviour already established.

\FloatBarrier
\clearpage
\subsection{EvoX Compiler: program transformation and performance}
\label{sec:evoxc-main}
\begin{figure}[H]
\centering
\includegraphics[width=\linewidth]{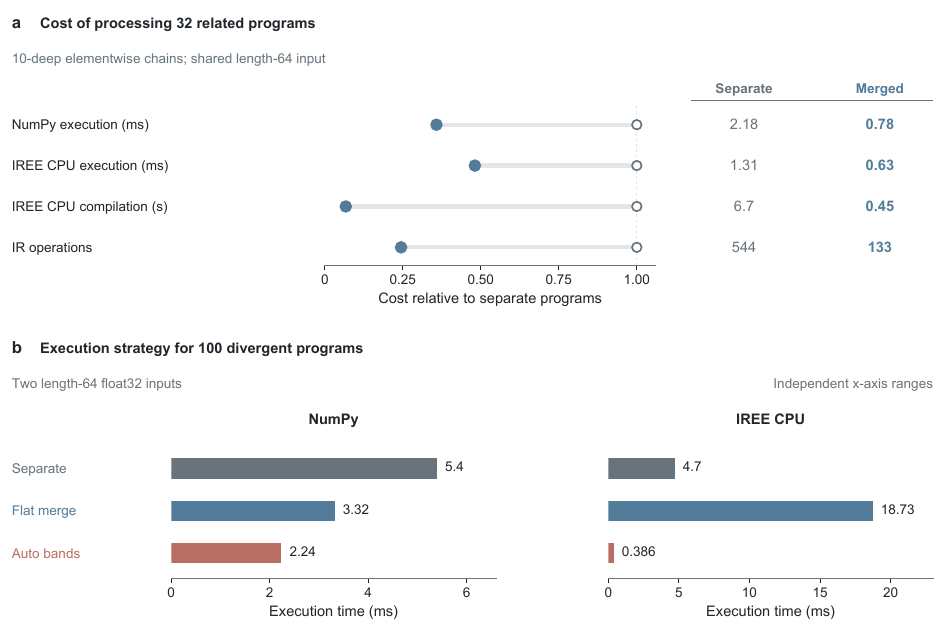}
\caption{\textbf{Program-merging strategy shapes execution and compilation costs in EvoX Compiler.}
\textbf{a}, For 32 related programs, filled points show merged cost relative to separate processing (open points, normalized to one); columns give absolute values in the indicated units. Compilation compares one merged IREE CPU compile with 32 separate compiles. Execution times are reported best-of-five values; compilation times and graph operation counts come from the same repository record.
\textbf{b}, Execution times for 100 divergent programs under separate execution, a flat merge and automatic banding, shown on separate linear scales for NumPy and IREE CPU. These are the repository's reported best times. Both panels use records retained at v0.1.0 (Table~\ref{tab:evidence-sources}, \protect\hyperlink{evidence:evoxcperf2026}{E19}); workloads and timing metadata are specified in Supplementary Section~\ref{sec:evoxc}.}
\label{fig:evoxc-results}
\end{figure}

The preceding studies follow continuity through inherited implementations and repeated checks. EvoX Compiler (\texttt{evoxc}), developed with Genesis, provides a complementary view of how changes within an implementation affect performance while remaining subject to executable interface requirements. Its EvoXIR pipeline supports IREE, XLA via PJRT and TVM (Table~\ref{tab:evidence-sources}, \protect\hyperlink{evidence:evoxc2026}{E16}); the recorded benchmarks examine how different program-merging strategies change execution and compilation costs (Fig.~\ref{fig:evoxc-results}).

For 32 related programs, merging reduces the operation count from 544 to 133 and IREE CPU compilation time from 6.7 to 0.45\,s. These reductions are accompanied by lower execution times, from 2.18 to 0.78\,ms on NumPy and from 1.31 to 0.63\,ms on IREE CPU. For the 100 divergent programs, however, the benefit depends on how merging is organized: a flat merge increases IREE CPU execution time from 4.7 to 18.73\,ms, whereas automatic banding reduces it to 0.386\,ms. With banding, NumPy execution time also falls from 5.4 to 2.24\,ms (Table~\ref{tab:evidence-sources}, \protect\hyperlink{evidence:evoxcperf2026}{E19}).

Performance is evaluated alongside executable requirements for the transformed programs and their backends. Tests compare compiled outputs with numerical references and check backend completion-event lifecycles (Table~\ref{tab:evidence-sources}, \protect\hyperlink{evidence:evoxcbatch2026}{E17}, \protect\hyperlink{evidence:evoxcsync2026}{E18}). Across the compiler and colocated HuggingFace integration suites, the release reports 7,511 passes with all extras and 7,086 in the compiler-free configuration, with zero failures in either. The retained tests therefore give later contributors concrete behavioural checks to use alongside the recorded costs when assessing further changes. Supplementary Section~\ref{sec:evoxc} gives the protocols and configuration.

\clearpage
\section{Discussion}

\subsection{Inheritance and revision in cumulative development}
The studies bring together two aspects of cumulative development: retaining accepted contributions and preserving their tested behaviour through later changes. Version ancestry makes the first visible, while repeated executable checks assess the second. In the compiler and renderer histories, these records meet at specific revisions, showing how a repair becomes part of the implementation and whether it survives further work. The renderer's regressions also show why evaluation must continue as the project grows. Sustaining development therefore involves both inheriting earlier contributions and reassessing their behaviour as new changes arrive.

What passes between episodes depends on the task. Compiler continuation preserves an implementation and its tests, so new agents can extend the same working system. MESA redevelopment changes the implementation itself; here, numerical references and precision checks provide continuity across the change of language. Despite this difference, both cases give later contributors concrete ways to assess what should be preserved. EvoX Compiler makes such expectations explicit at component boundaries, where graph transformations and backend operations have distinct executable requirements (Section~\ref{sec:evoxc-main}).

Because implementations, contracts and checks persist beyond individual episodes, successive agents can inspect, challenge and revise the same shared objects. Earlier work thus remains available for further evaluation after its original contributors have terminated. This continuity parallels Popper's emphasis on knowledge that stays open to criticism and further development beyond its authors~\cite{popper1972objective}.

\Needspace{12\baselineskip}
\subsection{Software evolution and operator roles}
The accumulated changes reside in the software and its development records. Although foundation model weights remain fixed during each run, the accepted project changes what subsequent agents can inspect, test and extend. Each new episode therefore begins with the consequences of earlier development embodied in its inherited repository.

Operators establish the conditions for this process through initial specifications, verification oracles, sandboxed tools and resource ceilings, and they trigger root phase transitions. Within those conditions, the recursive hierarchy performs task decomposition, code synthesis, debugging and local acceptance decisions autonomously, as recorded across the 1,019 compiler episodes and 272 MESA agents.

\subsection{Scope and planned component studies}
The four development studies document individual trajectories across formation, continuation and redevelopment, with one run for each experimental path. Observed delegation depths reached five in compiler formation, four and eight in the continuation lineages, and four in MESA redevelopment. These trajectories motivate replicated component studies to measure how the organization of work affects outcomes and how those outcomes vary across runs.

The planned studies use benchmarks of intermediate scale. A $2\times2$ factorial design crosses recursive delegation against flat coordination and transient episode resets against persistent sessions, separating spatial organization from session persistence. A complementary intervention holds executable code $\operatorname{Code}(v)$ fixed while varying inherited noncode records $\operatorname{Dev}(v)$, including directory contracts and provenance ledgers. Both designs will assess subsequent development through task completion, regressions on the inherited test suite and resource use.

\Needspace{7\baselineskip}
\subsection{Artifacts and software versions}
Repository snapshots identify the source of each reported result. Additional verification reports and reproduction entry points appear in Supplementary Section~\ref{sec:reproduction}. The accompanying software release is Genesis v0.14.1 (8 November 2026)~\cite{genesisrelease2026}; historical measurements retain their recorded versions and accounting conventions.

\Needspace{12\baselineskip}
\section{Conclusion}
\texttt{EvoX Genesis} organizes software development around a project that successive agents can inherit, evaluate and extend. In the compiler studies, local contributions assembled into a working system whose repairs survived a later revision, while a separate inherited compiler supported continued development across foundation model replacement. MESA redevelopment carried the question of continuity into a change of implementation language, and WebGL formation exposed repairs and regressions through repeated checks of API behaviour and rendered output. EvoX Compiler further connected implementation changes to performance and executable component requirements. Across these trajectories, the shared project retained both earlier work and concrete ways to assess what followed, making cumulative development observable beyond the lifetime of any individual agent.

\begingroup
\renewcommand*{\bibfont}{\small}
\setcounter{biburlnumpenalty}{100}
\setcounter{biburllcpenalty}{100}
\setcounter{biburlucpenalty}{100}
\printbibliography[title={References}]
\endgroup

\clearpage
\reportsupplementtrue
\colorlet{GenesisPurpleDark}{EvoXRed}

\title{Supplementary information for\\\textit{Persistent recursive worlds enable autonomous software evolution}}
\author{%
\textbf{Beichen Huang} \quad \textbf{Zhenyu Liang} \quad \textbf{Bowen Zheng} \quad \textbf{Ran Cheng}$^{*}$\\[0.30em]
Department of Data Science and Artificial Intelligence, The Hong Kong Polytechnic University\\
Hong Kong SAR, China\\[0.18em]
$^{*}$Correspondence: \href{mailto:ran-peter.cheng@polyu.edu.hk}{ran-peter.cheng@polyu.edu.hk}
}
\date{}

\hypersetup{pageanchor=false}
\setcounter{page}{1}
\gdef\thepage{S\arabic{page}}

\setcounter{section}{0}
\setcounter{subsection}{0}
\setcounter{subsubsection}{0}
\setcounter{figure}{0}
\setcounter{table}{0}
\setcounter{equation}{0}
\renewcommand{\thesection}{S\arabic{section}}
\renewcommand{\theHsection}{supp.\arabic{section}}
\renewcommand{\theHsubsection}{supp.\arabic{section}.\arabic{subsection}}
\renewcommand{\theHsubsubsection}{supp.\arabic{section}.\arabic{subsection}.\arabic{subsubsection}}
\renewcommand{\theHfigure}{supp.\arabic{figure}}
\renewcommand{\theHtable}{supp.\arabic{table}}
\renewcommand{\theHequation}{supp.\arabic{equation}}
\renewcommand{\figurename}{Supplementary Fig.}
\renewcommand{\thefigure}{S\arabic{figure}}
\renewcommand{\tablename}{Table}
\renewcommand{\thetable}{S\arabic{table}}
\maketitle
\thispagestyle{reportcover}

\makeatletter
\let\GenesisOriginalAddContentsLine\addcontentsline
\renewcommand{\addcontentsline}[3]{%
  \GenesisOriginalAddContentsLine{#1}{#2}{#3}%
  \def\GenesisContentsTarget{#1}%
  \def\GenesisMainToc{toc}%
  \ifx\GenesisContentsTarget\GenesisMainToc
    \GenesisOriginalAddContentsLine{suptoc}{#2}{#3}%
  \fi
}
\newcommand{\supplementarytableofcontents}{%
  \section*{Supplementary contents}%
  \@starttoc{suptoc}%
}
\makeatother
\begingroup\small\setcounter{tocdepth}{1}\setlength{\parskip}{3pt}
\makeatletter\renewcommand*{\l@section}{\@dottedtocline{1}{0em}{2.2em}}\makeatother
\supplementarytableofcontents\endgroup
\newpage

\section{Formal definition of persistent recursive worlds}

\subsection{Minimal model and persistent state}
The minimal model uses four formal objects: an accepted software version $v$, a directory path $p$ within the repository, a transient agent computation $A_i$ and an accepted software event. Together, these objects distinguish the project that persists from the episodes that modify it. Recursive organization, path context (\texttt{CONTEXT.md}), architectural constraints, validation results, reusable skills and provenance reside within the accepted version, where they remain available to subsequent agents.

\subsection{Local software worlds}
A local software world is defined by the coordinate pair
\begin{equation}
w=(v,p).
\label{eq:supp-local-world}
\end{equation}
The accepted version $v$ fixes the complete project state and its saved historical lineage, while the path $p$ specifies an agent's starting location, context, responsibility and modification scope. An agent can therefore inspect the complete project represented by $v$ while carrying out work anchored at $p$. The path identifies a local scope within that shared state, not a partial repository copy or a separate software state.

In the runtime, an accepted version corresponds to a validated commit, checkpoint or merge together with the versioned context files, constraints, test suites, skills and provenance records that later episodes inherit. This validation distinguishes accepted project states from arbitrary working trees. Isolated execution worktrees and rejected candidates remain outside the accepted version.

\subsection{Transient agency}
Given an episode objective $g_i$ in local world $(v,p)$, a transient agent synthesizes a candidate modification
\begin{equation}
\Delta_i = A_i\bigl((v,p),g_i\bigr).
\label{eq:supp-transient-agency}
\end{equation}
Within each bounded episode, managers and executors perform complementary roles. Root and intermediate managers decompose objectives, delegate subtasks and audit the evidence returned by their children; leaf executors synthesize and verify code modifications. Runtime labels such as codebase lead, codebase investigator and task scheduler specialize these roles. Both managers and executors maintain private execution state while working, then discard their conversational state when the episode terminates.

\subsection{Accepted events and verification}
A candidate modification enters the persistent version history exclusively through an accepted software event
\begin{equation}
(v,p)\longrightarrow(v',p').
\label{eq:supp-accepted-event}
\end{equation}
The event advances the accepted project from $v$ to $v'$, usually with $p'=p$. If the accepted change renames, moves or deletes a directory, its structural mapping instead determines $p'$. For delegated work, the responsible parent makes this acceptance decision after reviewing compiler diagnostics, test suites, scope checks and integration evidence; it may also reject the candidate or request refinement.

Rejection leaves $v$ unchanged and the candidate code outside the accepted lineage. Findings from the attempt can nevertheless inform subsequent work when they are committed to context files, tests, constraints or provenance records. In that case, the accepted version carries the findings forward after the agent's private conversational state has expired.

\Needspace{12\baselineskip}
\subsection{Recursive delegation}
Recursive delegation shifts the spatial locus of agency without altering the accepted version:
\begin{equation}
(v,p)\rightsquigarrow(v,q).
\label{eq:supp-recursive-delegation}
\end{equation}
A parent agent operating at $p$ instantiates a child at subpath $q$ within the same accepted version $v$. The child can delegate further within its scope, and leaf executors return verified code modifications up the hierarchy. Throughout this spatial decomposition, the accepted version remains fixed: delegation neither creates a new accepted state nor ratifies a child's output. The global version advances only when the parent accepts a returned contribution.

The two fundamental transitions are orthogonal:
\begin{center}
\begin{tabular}{@{}lll@{}}
\toprule
Transition operator & Version coordinate & Path coordinate \\
\midrule
Spatial recursive delegation $(v,p)\rightsquigarrow(v,q)$ & invariant ($v$) & projects from $p$ to $q$ \\
Temporal accepted event $(v,p)\rightarrow(v',p')$ & advances ($v \to v'$) & remains $p$ or maps to $p'$ \\
\bottomrule
\end{tabular}
\end{center}

\subsection{Persistent recursive worlds}
A persistent recursive world integrates two complementary forms of continuity:
\begin{enumerate}
\item \emph{temporal version inheritance}, in which validated consequences advance the persistent project from one version to the next; and
\item \emph{spatial recursive decomposition}, in which transient agents are instantiated across localized directory paths within the current accepted version~\cite{simon1962architecture}.
\end{enumerate}
Through these complementary transitions, subsequent episodes recover context from an accepted version and path. The persistent project records act as environmental traces~\cite{theraulaz1999stigmergy}: source code, \texttt{CONTEXT.md} contracts, constraints, validation suites, skills and provenance records.

\subsection{Reopening and structural changes}
Operations such as modify, expand, integrate, close, reopen, prune and regrow can be expressed through these accepted version transitions. For example, closing or reopening a module updates version history, while pruning and regrowth compose accepted events with path mappings. They therefore denote classes of transitions within the model; spatial delegation alone continues to leave the accepted version invariant.

\subsection{Scope of the evolutionary model}
The evolving object is the software project, including its source code, directory contracts, tests and versioned development records. Agents propose and evaluate changes within bounded episodes, and accepted commits determine what the next episodes inherit. Foundation model weights remain fixed during execution, while the inherited project changes through these accepted events.

\clearpage
\section{Evidence and measurement conventions}

\subsection{Evidence sources}
The compiler formation, compiler continuation and MESA migration packages provide task instructions, archived run summaries, repository versions and measurement tables. We read these records alongside the WebGL verification report and repair commits, the MESA interpolation golden tests, and the EvoX Compiler v0.1.0 source and release test summary. Results from these additional artifacts retain their committed records as sources; the sections below specify the test definitions and accounting scope for each.

\subsection{Repository versions}

Git commit IDs identify repository snapshots; external run archives and validation records are identified separately in each experiment.

\begingroup\small
\begin{table}[!htbp]
\centering
\caption{Repository versions for the four studies and the additional compiler artifact.}\label{tab:repository-commits}
\begin{tabular}{@{}lll@{}}

\toprule
Repository & Snapshot & Commit identifier \\
\midrule

jcc & DeepSeek formation & \href{https://github.com/EMI-Group/genesis-demo-jcc/commit/a8c116ec7ed3d59479ad3be8fed3e63384f50a87}{\texttt{a8c116ec7ed3}} \\
jcc & Initial GLM compiler & \href{https://github.com/EMI-Group/genesis-demo-jcc/commit/37216cfa254a40e40af48cd528743fd7f2d6737c}{\texttt{37216cfa254a}} \\
jcc & DeepSeek continuation & \href{https://github.com/EMI-Group/genesis-demo-jcc/commit/10077f65686102deef2c228e8a49b0569218ec48}{\texttt{10077f656861}} \\
jcc & GLM continuation & \href{https://github.com/EMI-Group/genesis-demo-jcc/commit/ab58a91460adb24e9349a554e02d4522d3ae2db4}{\texttt{ab58a91460ad}} \\
MESA & Fortran reference fork & \texttt{461dcba94f33} \\
mesa-rs & Migrated workspace & \href{https://github.com/EMI-Group/genesis-demo-mesa-rs/commit/2ad071e658f8d4de21aa9f0e906592c364e10023}{\texttt{2ad071e658f8}} \\
WebGL & Verification snapshot & \href{https://github.com/BillHuang2001/tbench-wasm/commit/67f620392b667714ba028f320d0e24f8ed32a16a}{\texttt{67f620392b66}} \\
WebGL & Tested renderer bundle & \href{https://github.com/BillHuang2001/tbench-wasm/commit/c4ffbff}{\texttt{c4ffbff}} \\
EvoX Compiler & v0.1.0 & \href{https://github.com/EMI-Group/evoxc/commit/2b289f034f6f8e16b6c9b55932dfed76cc54690e}{\texttt{2b289f034f6f}} \\
\bottomrule\end{tabular}
\end{table}\endgroup

Both continuation runs start from the GLM-generated Task~1 compiler at commit \texttt{37216cfa254a}. This shared starting point belongs to a separate lineage from the DeepSeek formation compiler at commit \texttt{a8c116ec7ed3}, which is therefore not used as the continuation baseline.

\subsection{Measurement and accounting conventions}
Unless a subsection states otherwise, an \emph{episode} is one row in \texttt{archive\_records}. Wall time is the top-level finish time minus the start time, whereas agent hours sum archived episode durations. Agent hours therefore exceed wall time when episodes overlap. To reconstruct parent and child trees, we exclude records whose \texttt{parent\_id} is absent from the archive, together with their descendants. Retention is assessed separately: a record is retained when its commit is an ancestor of the final repository, so this measure describes integration ancestry.

Physical repository lines count every line in tracked text files, including blank lines, comments and non-Rust files; file-type breakdowns distinguish Rust implementation files from other content. Token and cost quantities retain the accounting in the available run records, with no cross-experiment normalization when those conventions differ. Likewise, validation families retain their own denominators and meanings, and are reported separately rather than combined into a single score.

\Needspace{11\baselineskip}
\section{Formation: a C compiler from scratch}\label{sec:compiler-formation}

\subsection{Evidence source}
The main run record for compiler formation is \path{data/archive-deepseek-llvm.json}. Repository growth and composition were recomputed from the Git repository at final commit \href{https://github.com/EMI-Group/genesis-demo-jcc/commit/a8c116ec7ed3d59479ad3be8fed3e63384f50a87}{\texttt{a8c116ec7ed3}}; validation results come from the final root result and the committed test harness documentation. This DeepSeek run is separate from the older GLM~5.2 compiler used in Section~\ref{sec:compiler-continuation}.

\subsection{Task and initial specification}
The compiler experiment was one \texttt{Genesis} run from 2 August to 7 August 2026 (UTC) using DeepSeek V4 Flash. The task asked for an independent C compiler written in Rust for LLVM oriented toolchains and tests. It required a command-line interface compatible with Clang, standard object-file and linker integration, LLVM IR export and evaluation with the LLVM test suite. The compiler used a custom typed CIR with arena/index based storage rather than LLVM IR, and direct use or translation of Clang/LLVM source was prohibited. C11 was the main language target, C23 was a stretch goal, x86 and x86-64 were required back ends and AArch64 was optional. These specifications define the design, toolchain and testing requirements for formation from a repository with no compiler implementation.

\subsection{Initial repository and run phases}
The first root session began at commit \texttt{41e087ce90f3}, which contained only \texttt{.gitignore} and \texttt{genesis.toml}, with no compiler implementation. Delegated agents developed their contributions in separate Git worktrees. The second root session subsequently inherited the first session's final commit, \texttt{04d64b89d504}, together with a handoff summary covering completed work, validation results, known gaps and next steps. The two sessions thus form consecutive initialization and optimization phases of the same run.

\subsection{Operator specification}
The root objective specified the implementation language, high level compiler organization, typed intermediate representation, command-line compatibility, target architectures and test sources. Genesis generated the compiler code and concrete module implementations under this operator specification.

\subsection{Evaluation targets}
The language targets were C11 and selected C23 extensions. The test suites and integration workloads below evaluate compiler behaviour within this scope.

\subsection{External validation sources}
LLVM (upstream: \url{https://github.com/llvm/llvm-project}) supplies the toolchain workflows and test programs. Compatibility is evaluated at the command-line and toolchain interfaces, while the compiler retains its own typed CIR and supports LLVM IR export. Verification then combines external programs from c-testsuite, the LLVM test suite, LZ4 and SQLite with Csmith programs generated by the committed harness. Each family retains its testing unit and coverage in the reported results.

\Needspace{9\baselineskip}
\subsection{Run configuration and archive structure}
The run comprised two sequential root sessions. Its archive contains 1,019 episode records, whereas the top-level \texttt{agent\_count} reports 1,065 spawned agents; archive-based organization statistics therefore describe the records actually present. Four of these records reference missing parent IDs and have no archived descendants. Excluding them leaves 1,015 records for the reconstructed parent and child tree. Token and cost totals instead use the top-level usage record. The archived \texttt{objective} fields retain task instructions, but complete system prompts, message histories and tool outputs are unavailable.

\subsection{Developmental lineage and complete results}
Run totals, archived episodes, records in the reconstructed parent and child tree, Git history, repository line counts and test results refer to different sets of records, so they are reported separately.

\begin{figure}[H]
\centering
\includegraphics[width=\linewidth]{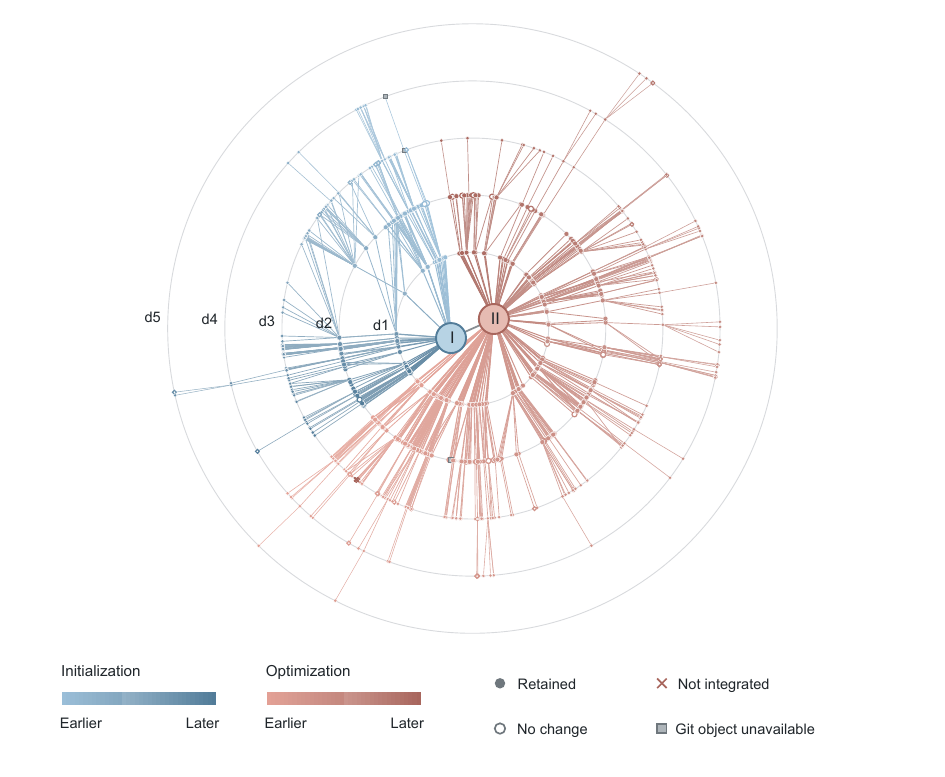}
\caption{\textbf{DeepSeek compiler development tree.} The radial tree shows the two root sessions and all 1,015 records used in the reconstructed parent and child tree after four records with missing parents were excluded. Radial distance shows delegation depth. Blue and coral show initialization and optimization; shade shows completion order within each phase. Symbols mark retained, no change, not integrated and Git object unavailable outcomes.}
\label{fig:supp-compiler-lineage}
\end{figure}

\begingroup
\small
\setlength{\tabcolsep}{3pt}
\begin{table}[!htbp]
\centering
\caption{Compiler archive and integration counts.}\label{tab:compiler-lineage-accounting}
\begin{tabular}{@{}>{\raggedright\arraybackslash}p{0.54\linewidth}r@{}}

\toprule
Quantity & Value \\
\midrule

Raw archived records & 1,019 \\
Top level \texttt{agent\_count} field & 1,065 \\
Direct missing parent records & 4 \\
Excluded descendants & 0 \\
Records in reconstructed tree & 1,015 \\
Initialization records & 312 \\
Optimization records & 703 \\
Retained & 929 \\
Retention among reconstructed records & 91.53\% \\
No change & 78 \\
Not integrated & 5 \\
Git object unavailable & 3 \\
First-parent commits & 327 \\
Maximum observed delegation depth & 5 \\
Peak active episodes & 29 \\
\bottomrule
\end{tabular}
\end{table}
\endgroup

Retention locates a contribution in the Git history leading to the final repository, while the validation suites assess the functionality of that repository. These measurements therefore describe complementary aspects of the run. Within the archived history, the deepest record reached delegation depth 5 against a configured maximum of 8.

\subsubsection{Resource use and repository size}
The run lasted 123.402~h of wall time. Because multiple episodes overlapped, their summed durations reached 666.385 agent hours. Cached input accounted for 97.382\% of input tokens.

\begingroup
\small
\setlength{\tabcolsep}{3pt}
\begin{table}[!htbp]
\centering
\caption{Compiler formation resource use.}\label{tab:compiler-resources}
\begin{tabular}{@{}>{\raggedright\arraybackslash}p{0.50\linewidth}r@{}}

\toprule
Metric & Observed value \\
\midrule

Elapsed wall time & 123.402 h \\
Raw agent hours & 666.385 h \\
Median episode duration & 12.76 min \\
Input tokens & 4,134,593,954 \\
Cached input tokens & 4,026,336,896 \\
Uncached input tokens & 108,257,058 \\
Cached input/input & 97.382\% \\
Output tokens & 64,092,688 \\
Total tokens & 4,198,686,642 \\
Logged cost & US\$44.3760 \\
\bottomrule
\end{tabular}
\end{table}
\endgroup

\begingroup
\small
\setlength{\tabcolsep}{3pt}
\begin{table}[!htbp]
\centering
\caption{Compiler formation resource use by phase.}\label{tab:compiler-phases}
\begin{tabular}{@{}>{\raggedright\arraybackslash}p{0.20\linewidth}rrrrr@{}}

\toprule
Phase & Records & Wall time (h) & Agent hours & Total tokens & Cost (US\$) \\
\midrule

Initialization & 312 & 23.905 & 163.925 & 1,052,907,912 & 13.5715 \\
Optimization & 707 & 99.497 & 502.460 & 3,145,778,730 & 30.8045 \\
Total & 1,019 & 123.402 & 666.385 & 4,198,686,642 & 44.3760 \\
\bottomrule
\end{tabular}
\end{table}
\endgroup

Phase totals use the two nonoverlapping root usage records. Since each parent record already includes usage from its descendants, summing per-episode usage for billing would count that usage again.

Physical line counts include blank lines and comments in every tracked text file. The final repository contains 248,989 physical lines across 750 text files and no tracked binary files; 219,676 lines are in 354 Rust source files. The remaining lines include documentation, tests and validation material. The 248,989 total measures the complete tracked text repository.

\begin{table}[!htbp]
\centering
\begin{minipage}[t]{\linewidth}
\caption{Compiler physical lines at handoff and completion.}\label{tab:compiler-composition-stages}
\centering\small
\begin{tabular}{@{}lrr@{}}
\toprule
Functional category & Handoff & Final \\
\midrule
Front end & 29,102 & 73,842 \\
IR and optimization & 32,862 & 70,926 \\
Back end & 31,034 & 76,021 \\
Supporting code & 20,782 & 28,200 \\
Total & 113,780 & 248,989 \\
\bottomrule
\end{tabular}
\end{minipage}\par\medskip
\begin{minipage}[t]{\linewidth}
\caption{Classified non-root compiler work allocation.}\label{tab:compiler-work-allocation}
\centering\small
\begin{tabular}{@{}lrr@{}}
\toprule
Work category & Episodes & Agent-hours \\
\midrule
Front end & 324 & 197.259 \\
IR and optimization & 248 & 118.982 \\
Back end & 248 & 162.721 \\
Integration / validation & 185 & 57.194 \\
Cross-cutting & 8 & 5.116 \\
Total & 1,013 & 541.271 \\
\bottomrule
\end{tabular}
\end{minipage}
\end{table}

Work allocation is classified from Git diffs for 932 episodes, objective text for 78 no-change episodes and objective text for three episodes whose Git objects are unavailable. The classified set excludes the two roots and four records with missing parents. Its category durations and total retain the source record's three-decimal precision. By contrast, the raw episode durations in Table~\ref{tab:compiler-phases} cover the complete archive.

\begin{figure}[H]
\centering
\includegraphics[width=\linewidth]{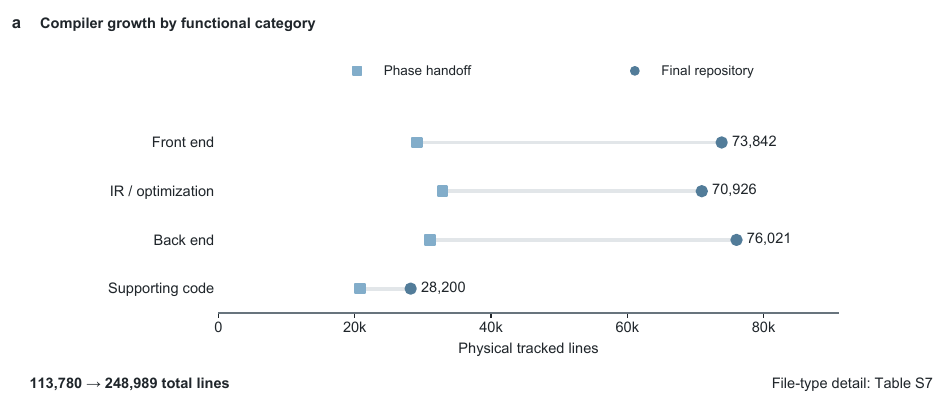}
\caption{\textbf{Compiler growth by functional category.} Squares show physical tracked lines at the phase handoff; circles show the final repository. Totals grow from 113,780 to 248,989 lines. Counts include comments, blank lines and other tracked text. Table~\ref{tab:compiler-file-types} provides the file-type breakdown.}
\label{fig:supp-compiler-composition}
\end{figure}

\begingroup
\scriptsize
\setlength{\tabcolsep}{3pt}
\begin{table}[!htbp]
\centering
\caption{Final compiler repository lines by tracked text file type.}\label{tab:compiler-file-types}
\begin{tabular}{@{}>{\raggedright\arraybackslash}p{0.31\linewidth}rrr@{}}

\toprule
File type & Files & Physical lines & Share \\
\midrule

Rust & 354 & 219,676 & 88.23\% \\
Markdown & 77 & 17,005 & 6.83\% \\
C source & 130 & 5,036 & 2.02\% \\
Python & 14 & 3,265 & 1.31\% \\
C header & 49 & 3,163 & 1.27\% \\
Expected output & 106 & 271 & 0.11\% \\
Shell & 3 & 195 & 0.08\% \\
Cargo.toml & 9 & 132 & 0.05\% \\
Makefile & 1 & 101 & 0.04\% \\
Cargo.lock & 1 & 66 & 0.03\% \\
TOML & 2 & 42 & 0.02\% \\
YAML & 1 & 25 & 0.01\% \\
\texttt{.gitignore} & 2 & 10 & 0.00\% \\
Info & 1 & 2 & 0.00\% \\
Total & 750 & 248,989 & 100.00\% \\
\bottomrule
\end{tabular}
\end{table}
\endgroup

\begingroup
\small
\setlength{\tabcolsep}{3pt}
\begin{table}[!htbp]
\centering
\caption{Final compiler test results and limits.}\label{tab:compiler-validation}
\begin{tabular}{@{}>{\raggedright\arraybackslash}p{0.18\linewidth}>{\raggedright\arraybackslash}p{0.17\linewidth}>{\raggedright\arraybackslash}p{0.18\linewidth}r >{\raggedright\arraybackslash}p{0.30\linewidth}@{}}

\toprule
Test & Measure & Result & Rate & Notes \\
\midrule

LLVM test suite & Passing cases & 32/36 & 88.9\% & Four reported cases did not pass \\
c-testsuite & Passing tests & 220/220 & 100.0\% & Complete reported c-testsuite set \\
Csmith & Fixed generated programs & 93/100 & 93.0\% & 93 passes; seven programs time out under both jcc and Clang \\
LZ4 & Essential checks & 8/8 & 100.0\% & All reported essential checks passed \\
SQLite & Test stages & 2/2 & 100.0\% & Compile/link and deterministic SQL checks; not the upstream suite \\
Rust unit tests & Workspace count & 2,904 passed & N/A & One intentional ignore; no fixed external denominator \\
Internal corpus & Compiler cases & 106/106 & 100.0\% & 86 compile and run and 20 compile fail cases \\
\bottomrule
\end{tabular}
\end{table}
\endgroup

Each family retains its verification unit. SQLite measures shell compilation and deterministic queries; Csmith evaluates 100 generated seeds, with 93 passes and seven cases that time out under both compilers.

\begingroup
\small
\setlength{\tabcolsep}{3pt}
\begin{table}[!htbp]
\centering
\caption{Compiler task and run settings.}\label{tab:compiler-settings}
\begin{tabular}{@{}>{\raggedright\arraybackslash}p{0.31\linewidth}>{\raggedright\arraybackslash}p{0.64\linewidth}@{}}

\toprule
Item & Setting \\
\midrule

Run setup & \texttt{Genesis}; two sequential root sessions; per agent Git worktrees \\
Model & DeepSeek V4 Flash \\
Archived model identifier & \texttt{deepseek:deepseek-v4-flash}; \texttt{model\_id=deepseek flash} \\
Reasoning effort & \texttt{xhigh} \\
Context compression threshold & 150,000 tokens \\
Initial codebase & \texttt{.gitignore} and \texttt{genesis.toml}; no compiler implementation \\
Root task inputs & Phase I compiler blueprint; Phase II continuation with handoff summary \\
Turn limits & 2,048 root turns; 128 turns per nonroot episode \\
Delegation and retry limits & Maximum depth 8; maximum retries 15 \\
External test sources & c-testsuite, LLVM test suite, LZ4 and SQLite \\
Generated program tests & Csmith through the committed harness \\
Rust toolchain & Rust stable; edition 2024; minimum Rust version 1.85; rustfmt and clippy \\
Compiler target & C11 primary and C23 stretch; x86 and x86-64 mandatory; AArch64 optional \\
\bottomrule
\end{tabular}
\end{table}
\endgroup

\begingroup
\scriptsize
\setlength{\tabcolsep}{3pt}
\begin{table}[!htbp]
\centering
\caption{Machine and software environment for the compiler experiments.}\label{tab:compiler-environment}
\begin{tabular}{@{}>{\raggedright\arraybackslash}p{0.28\linewidth}>{\raggedright\arraybackslash}p{0.66\linewidth}@{}}

\toprule
Item & Recorded value \\
\midrule

CPU & AMD Ryzen 7 PRO 6850HS with Radeon Graphics \\
CPU layout & x86-64; one socket; 8 physical cores; 2 threads per core; 16 online logical CPUs \\
Memory & 64 GB \\
CPU frequency & 403.7300--4787.0820 MHz; frequency boost enabled \\
Cache & 256 KiB L1d, 256 KiB L1i, 4 MiB L2 and 16 MiB L3 in aggregate \\
NUMA & One node containing CPUs 0--15 \\
Container and host & ArchLinux container; NixOS built Linux 6.18.39 host kernel \\
Filesystem & ZFS with ZFS clone worktrees \\
Tool isolation & \texttt{systemd-run} and cgroups for every agent tool call \\
Missing reproduction fields & Container digest, package lock, per tool cgroup limits, exact Rust/linker revisions and host load traces \\
\bottomrule
\end{tabular}
\end{table}
\endgroup

The compiler formation and continuation runs used the same dedicated machine. Within the continuation study, concurrency, CPU frequency, agent counts and token use varied between the two paths.

\subsection{Differential testing and retained repairs}
\label{sec:csmith-repairs}
The committed Csmith record follows the same 100 generated programs through three compiler revisions (Table~\ref{tab:evidence-sources}, \protect\hyperlink{evidence:jcccsmith2026}{E01}). Runs A875 and A965 reuse the existing files, and the latter recheck verifies byte identity for all 100 programs. Both jcc and Clang 22.1.8 run at \texttt{-O1}, with compilation limited to 600 seconds and execution to 300 seconds, a 32~GiB address-space limit and eight parallel workers. A successful program matches Clang's output checksum. Table~\ref{tab:csmith-repair-stages} consequently retains every generated seed in its total while recording how its outcome changes.

\begin{table}[!htbp]
\centering\small
\caption{Recorded outcomes for a fixed set of 100 Csmith programs.}\label{tab:csmith-repair-stages}
\begin{tabular}{@{}llrrrr@{}}
\toprule
Run & Compiler commit & Pass & Compile error & Miscompile & Both timeout \\
\midrule
A850 & \texttt{61bdf2bc2} & 61 & 29 & 6 & 4 \\
A875 & \texttt{60eed7a3c} & 93 & 0 & 0 & 7 \\
A965 & \texttt{48176a41e} & 93 & 0 & 0 & 7 \\
\bottomrule
\end{tabular}
\end{table}

Fixes to unnamed bit-field initialization, bit-field unit closure and packed accesses converted all six miscompiles (seeds 27, 30, 35, 56, 87 and 93) into passes. Of the 29 register-allocation compilation errors, 26 also became passes. The remaining three programs (20, 66 and 88) compiled successfully but timed out under both compilers, joining the four existing timeout cases. All 61 original passes were retained through these repairs. The subsequent A965 recheck preserved the complete A875 classification, including the same seven timeout seeds: 20, 22, 60, 66, 73, 81 and 88. Figure~\ref{fig:compiler-development}b uses these outcome classes, with timings labelled slow in the source grouped with successful passes.

\subsection{Archive completeness}
The archive status is \texttt{completed}. Its raw count retains four records with missing parent IDs, while the reconstructed parent and child tree excludes them. The repository and test measurements can be traced through the available archive; exact execution replay would additionally require the missing provider, message and environment details.

\section{Continuation: a compiler across foundation models}\label{sec:compiler-continuation}

\subsection{Shared starting world}
The continuation experiment begins with a GLM~5.2 compiler history separate from the DeepSeek formation run in Section~\ref{sec:compiler-formation}. Initial GLM development produced commit \texttt{37216cfa254a}, which both continuation branches then inherited. One branch continued with GLM~5.2 and the other with DeepSeek V4 Flash. The three stages thus share one ancestral GLM snapshot as the baseline for examining subsequent development.

Both continuation runs received the same completed GLM repository, saved project context, user instruction, test families and recorded controller limits. The instruction was: \emph{``Continue and finish all remaining work, achieve 100\% pass rate, excluding csmith as it is not installed.''} Agent count, concurrency, token use, wall time and code growth were not fixed in advance.

\subsection{Evaluation protocol}
Both continuations used Rust workspace tests, LLVM SingleSource, c-testsuite, LZ4 and SQLite at \texttt{-O0}. LLVM pass fractions use the retained case manifest of each snapshot.

\subsection{Evaluation changes during continuation}
\label{sec:continuation-evaluation-changes}
The initial evaluation collected 1,870 cases and reported 1,558 passes, 275 run failures and 37 compilation failures. An audit traced a comparison defect to a trailing newline that prevented removal of the reference output's final \texttt{exit N} marker. Re-executing the 275 run failures with corrected handling classified 190 as passing programs rejected by the runner, leaving 85 residual run failures (Table~\ref{tab:evidence-sources}, \protect\hyperlink{evidence:jccllvmaudit2026}{E02}). Further runner changes addressed command-line arguments, stdout/stderr comparison and per-test build flags.

DeepSeek's committed test notes document three collection changes (Table~\ref{tab:evidence-sources}, \protect\hyperlink{evidence:jccdsprotocol2026}{E03}). Five exclusions first reduced 1,870 to 1,865 cases: two helper programs, an AArch64 case, a vector-shuffle case and a variadic-pack case. A later batch removed 31 hash-reference cases and another AArch64 case, producing 1,833. The final batch removed seven arbitrary-precision integer cases, four integer-reference cases and two floating-point-reference cases, producing 1,820. These collection changes define the denominator of the final 1,820/1,820 result at commit \texttt{10077f656861}.

GLM's final runner applies its own collection rules, excluding the \texttt{AArch64} and \texttt{Integer} directories alongside seven named files and its main-function filters (Table~\ref{tab:evidence-sources}, \protect\hyperlink{evidence:jccglmprotocol2026}{E04}). The resulting retained suite gives 1,445/1,448 at commit \texttt{ab58a91460ad}. Each branch's fraction therefore describes the cases retained by its final runner. Alongside these manifest changes, workspace tests increase from 1,136 to 1,226 and 1,350, respectively, while c-testsuite remains 220/220 in all three snapshots. The figure data preserve both final filter sets and the audit's classification.

\subsection{Run settings, lineage and results}
Each continuation path comprises one completed run.

Both continuations used the same recorded limits for depth, retries, turns and context compression, and both ran on the machine in Table~\ref{tab:compiler-environment}.

\begingroup
\small
\setlength{\tabcolsep}{3pt}
\begin{table}[!htbp]
\centering
\caption{Compiler continuation starting points and run settings.}\label{tab:continuation-settings}
\begin{tabular}{@{}>{\raggedright\arraybackslash}p{0.23\linewidth}>{\raggedright\arraybackslash}p{0.23\linewidth}>{\raggedright\arraybackslash}p{0.23\linewidth}>{\raggedright\arraybackslash}p{0.23\linewidth}@{}}

\toprule
Setting & Initial development & GLM continuation & DeepSeek continuation \\
\midrule

Model path & New project $\rightarrow$ GLM~5.2 & GLM compiler $\rightarrow$ GLM~5.2 & Same GLM compiler $\rightarrow$ DeepSeek V4 Flash \\
Starting state & Empty tracked project & Completed compiler at \texttt{37216cfa254a} & Same completed compiler at \texttt{37216cfa254a} \\
Task & Build new compiler & Continue existing compiler & Continue existing compiler \\
Root role & Not recorded & Manager & Manager \\
Depth and retry limits & 8; 15 & 8; 15 & 8; 15 \\
Turn limits & 2,048 root; 128 delegated & 2,048 root; 128 delegated & 2,048 root; 128 delegated \\
Compression threshold & 150,000 tokens & 150,000 tokens & 150,000 tokens \\
Test families & Rust, LLVM, c-testsuite, LZ4, SQLite & Same families & Same families \\
Csmith & Requested; unavailable & Excluded & Excluded \\
Runs & 1 & 1 & 1 \\
Shared fixed budget & Not specified & Not specified & Not specified \\
\bottomrule
\end{tabular}
\end{table}
\endgroup

\begingroup
\small
\setlength{\tabcolsep}{3pt}
\begin{table}[!htbp]
\centering
\caption{Run summary for compiler continuation.}\label{tab:continuation-summary}
\begin{tabular*}{\linewidth}{@{\extracolsep{\fill}}>{\raggedright\arraybackslash}p{0.37\linewidth}rrr@{}}

\toprule
Metric & \shortstack[r]{Initial\\GLM} & \shortstack[r]{GLM\\continuation} & \shortstack[r]{DeepSeek\\continuation} \\
\midrule

Elapsed time (h) & 136.56 & 21.99 & 17.10 \\
Spawned agents & 562 & 98 & 178 \\
Archived records & 504 & 97 & 168 \\
Archive coverage & 89.7\% & 99.0\% & 94.4\% \\
First-parent commits & 619 & 88 & 31 \\
Maximum observed depth & 5 & 4 & 8 \\
Peak active agents & 21 & 9 & 19 \\
Mean active agents & 2.86 & 2.94 & 5.70 \\
Summed agent hours & 390.57 & 64.65 & 97.57 \\
Median duration (min) & 14.4 & 16.1 & 17.0 \\
90th percentile duration (min) & 83.2 & 59.8 & 70.2 \\
Code changing records & 479 & 90 & 160 \\
No change records & 25 & 7 & 8 \\
\bottomrule
\end{tabular*}
\end{table}
\endgroup

DeepSeek completed continuation in less elapsed time, but used more agents, greater concurrency, more archived agent hours and more tokens. Table~\ref{tab:continuation-summary} therefore reports these resource dimensions separately, so the shorter elapsed time can be read alongside the resources consumed.

\begingroup
\small
\setlength{\tabcolsep}{3pt}
\begin{table}[!htbp]
\centering
\caption{Archive coverage for compiler continuation.}\label{tab:continuation-coverage}
\begin{tabular}{@{}>{\raggedright\arraybackslash}p{0.26\linewidth}rrr>{\raggedright\arraybackslash}p{0.27\linewidth}@{}}

\toprule
Stage & Spawned & Archived & Coverage & Missing parent issue \\
\midrule

Initial development & 562 & 504 & 89.7\% & Six absent parent IDs referenced by 15 records \\
GLM continuation & 98 & 97 & 99.0\% & Two records reference absent parent 562 \\
DeepSeek continuation & 178 & 168 & 94.4\% & No missing parent reference among archived records \\
\bottomrule
\end{tabular}
\end{table}
\endgroup

Lineage, duration and concurrency statistics are reconstructed from the archived records, with coverage and missing parent links listed in Table~\ref{tab:continuation-coverage}.

\begin{figure}[H]
\centering
\includegraphics[width=\linewidth]{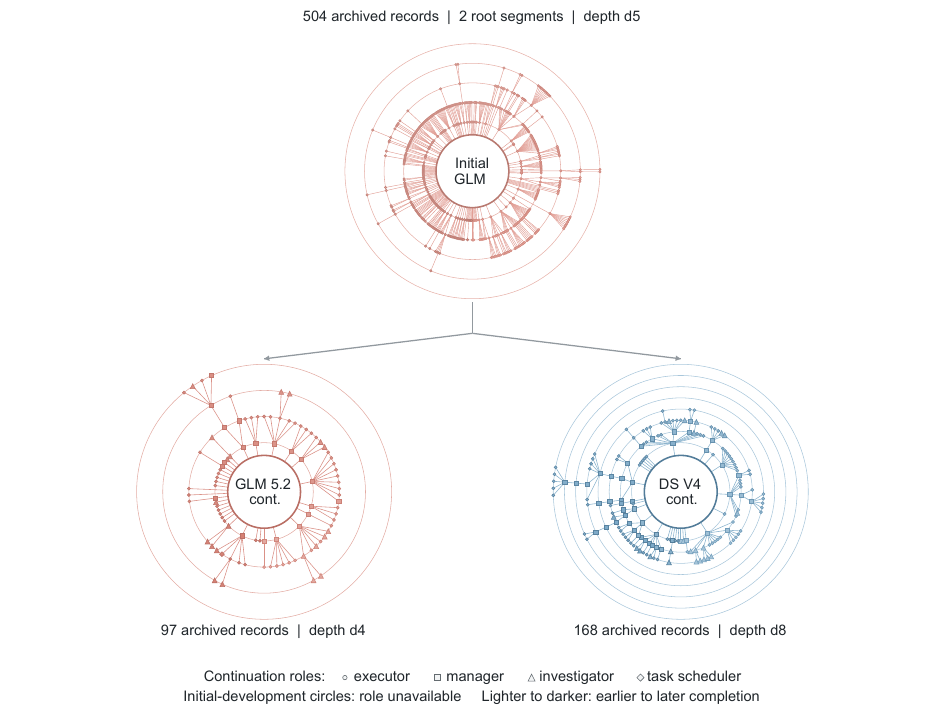}
\caption{\textbf{Delegation trees for compiler continuation.} The upper tree shows the 504 archived records from initial GLM development, including two root segments and depth to d5; agent roles were not recorded for this stage. The lower trees show the 97 archived GLM~5.2 continuation records and 168 archived DeepSeek V4 Flash continuation records. Distance from the centre shows delegation depth, node shape shows the recorded role where available and colour shade shows completion order.}
\label{fig:supp-continuation-lineages}
\end{figure}

\begin{figure}[H]
\centering
\includegraphics[width=\linewidth]{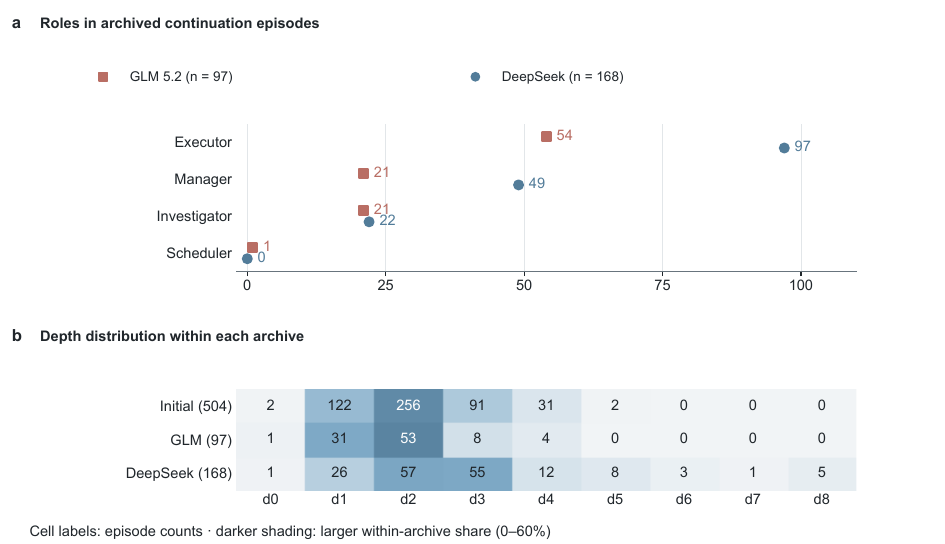}
\caption{\textbf{Agent roles and delegation depth in compiler continuation.} \textbf{a}, Recorded role counts in the 97 archived GLM and 168 archived DeepSeek continuation episodes. \textbf{b}, Each cell gives an episode count; shading shows its share of the corresponding archive, on a common 0 to 60\% scale. Initial development contains 504 archived episodes. Roles were recorded for the continuation runs.}
\label{fig:supp-continuation-organization}
\end{figure}

\subsubsection{Resources, repository growth and validation}
Both continuation runs achieved high prompt cache reuse. Resource consumption is reported per lineage under the shared controller limits and independently realized budgets.

\begingroup
\small
\setlength{\tabcolsep}{3pt}
\begin{table}[!htbp]
\centering
\caption{Token use and cost in compiler continuation.}\label{tab:continuation-tokens}
\begin{tabular}{@{}lrrr@{}}

\toprule
Metric & Initial GLM & GLM continuation & DeepSeek continuation \\
\midrule

Input tokens & 2,245,871,926 & 543,563,883 & 902,775,744 \\
Cached input tokens & 2,163,779,584 & 534,150,656 & 884,375,552 \\
Fresh input tokens & 82,092,342 & 9,413,227 & 18,400,192 \\
Cached input/input & 96.34\% & 98.27\% & 97.96\% \\
Output tokens & 19,334,886 & 3,702,697 & 8,704,733 \\
Total tokens & 2,265,206,812 & 547,266,580 & 911,480,477 \\
Input cost (US\$) & 677.511916 & 152.057688$^{*}$ & 5.052314 \\
Output cost (US\$) & 85.073659 & 16.291867$^{*}$ & 2.437395 \\
Total cost (US\$) & 762.585575 & 168.349555$^{*}$ & 7.489709 \\
\bottomrule
\multicolumn{4}{@{}l}{\footnotesize $^{*}$Reconstructed cost; other cost entries are archive-recorded.}\\
\end{tabular}
\end{table}
\endgroup

The GLM continuation cost is reconstructed from its token counts and the rate schedule matching the initial GLM bill. Initial-development and DeepSeek costs are recorded in their archives.

\begingroup
\small
\setlength{\tabcolsep}{3pt}
\begin{table}[!htbp]
\centering
\caption{Tests reported at the end of compiler continuation.}\label{tab:continuation-validation}
\begin{tabular}{@{}>{\raggedright\arraybackslash}p{0.25\linewidth}>{\raggedright\arraybackslash}p{0.22\linewidth}>{\raggedright\arraybackslash}p{0.22\linewidth}>{\raggedright\arraybackslash}p{0.22\linewidth}@{}}

\toprule
Test target & Initial development & GLM continuation & DeepSeek continuation \\
\midrule

Rust unit tests & 1,136 & 1,226 ($+90$) & 1,350 ($+214$) \\
LLVM SingleSource & 1,558/1,870 (83.3\%) & 1,445/1,448 (99.79\%) & 1,820/1,820 (100\%) \\
c-testsuite & 220/220 & 220/220 & 220/220 \\
LZ4 & 4/4 files & 4/4 files & 4/4 files \\
SQLite at \texttt{-O0} & Basic run passed & Compiled & Compiled, linked and ran \\
Csmith & Unavailable & Excluded & Excluded \\
\bottomrule
\end{tabular}
\end{table}
\endgroup

Because the LLVM case lists change between snapshots, each pass fraction applies to its own saved manifest. The archive identifies three nonpassing GLM continuation cases, although it does not retain per-case diagnostics. The c-testsuite, LZ4 and SQLite outcomes likewise come from saved completion records; these suites were not rerun while preparing this Supplementary Information.

\begingroup
\small
\setlength{\tabcolsep}{3pt}
\begin{table}[!htbp]
\centering
\caption{Physical line counts for the compiler continuation snapshots.}\label{tab:continuation-lines}
\begin{tabular}{@{}>{\raggedright\arraybackslash}p{0.29\linewidth}rrr@{}}

\toprule
Included file type & Task~1 & GLM 5.2 & DeepSeek V4 Flash \\
\midrule

Rust & 94,253 & 104,264 & 117,409 \\
Cargo.toml & 202 & 202 & 202 \\
Markdown & 10,325 & 11,876 & 14,718 \\
Shell & 640 & 649 & 825 \\
Total & 105,420 & 116,991 & 133,154 \\
\bottomrule
\end{tabular}
\end{table}
\endgroup

The initial-development total in Table~\ref{tab:continuation-lines} belongs to the GLM compiler history and includes tracked Rust, \texttt{Cargo.toml}, Markdown and shell files. In contrast, the 248,989-line formation total in Section~\ref{sec:compiler-formation} covers all tracked text files in the separate DeepSeek repository. The totals therefore retain both the compiler lineage and the file scope specified in their respective tables.

\begin{figure}[H]
\centering
\includegraphics[width=\linewidth]{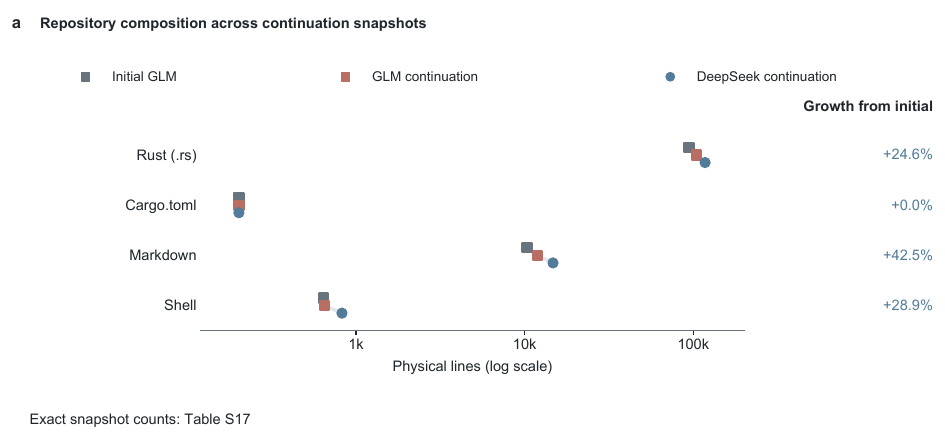}
\caption{\textbf{Repository composition across continuation snapshots.} Symbols identify initial GLM development, GLM continuation and DeepSeek continuation. Physical line counts include comments and blank lines in tracked Rust, Cargo.toml, Markdown and shell files. The right column gives the percentage increase from initial GLM to DeepSeek continuation. Exact counts appear in Table~\ref{tab:continuation-lines}.}
\label{fig:supp-continuation-lines}
\end{figure}

Both GLM~5.2 and DeepSeek V4 Flash extended the inherited compiler through newly instantiated agents. The shared repository thus supported further development after agent turnover and foundation model replacement. The final section specifies planned component interventions to examine how particular inherited records contribute to that process.

\clearpage
\section{Redevelopment: MESA from Fortran to Rust}\label{sec:mesa}

\subsection{Starting system and run settings}
The experiment used a lightly modified MESA fork (upstream: \url{https://github.com/MESAHub/mesa}) at commit \path{461dcba94f33a0dd991db129d406a877fcc9dfdc} as an immutable reference. The Rust output is archived at commit \path{2ad071e658f8d4de21aa9f0e906592c364e10023}. The run used DeepSeek V4 Flash with \texttt{xhigh} reasoning effort and a context compression threshold of 150,000 tokens.

Redevelopment comprised one run, with a second root-level objective beginning at 31.720~h after a handoff from the first. Resource totals use the top-level archive record, since parent records already include usage from their descendants.

The root objective requested a broad Rust rewrite with API compatibility, testing and performance optimization. The implemented scope comprises selected functionality across the 13 mapped foundation and physics modules, evaluated with the workloads below.

\begingroup
\small
\setlength{\tabcolsep}{3pt}
\begin{table}[!htbp]
\centering
\caption{MESA to Rust run and timing settings.}\label{tab:migration-settings}
\begin{tabular}{@{}>{\raggedright\arraybackslash}p{0.31\linewidth}>{\raggedright\arraybackslash}p{0.64\linewidth}@{}}

\toprule
Quantity & Value \\
\midrule

Model & DeepSeek V4 Flash \\
Model settings & \texttt{xhigh} reasoning; compression threshold of 150,000 tokens \\
Recorded agent roles & Codebase lead, manager, executor and codebase investigator \\
Controller limits & Maximum depth 8; retries 15; 2,048 root turns and 128 child turns \\
Source baseline & Modified MESA fork, commit \texttt{461dcba94f33} \\
Timing host & Shared Intel Xeon Platinum 8336C system; 64 physical cores / 128 hardware threads; benchmark processes pinned to CPUs 0 to 3 \\
Timing runs & 25 direct binary runs per workload after warmup; separate burn proxy check of 40 runs \\
Fortran build & gfortran 12.2.0; \texttt{-O3 -march=native -ffp-contract=fast -std=f2008} \\
Rust build & \texttt{opt-level=3}; fat LTO; \texttt{codegen-units=1}; \texttt{panic=abort} \\
\bottomrule
\end{tabular}
\end{table}
\endgroup

\begingroup
\small
\setlength{\tabcolsep}{3pt}
\begin{table}[!htbp]
\centering
\caption{MESA to Rust resource use and archive coverage.}\label{tab:migration-resources}
\begin{tabular}{@{}>{\raggedright\arraybackslash}p{0.55\linewidth}r@{}}

\toprule
Quantity & Recorded value \\
\midrule

Experiment interval & 3--5 August 2026 (UTC) \\
Elapsed wall time & 33.219 h \\
Root agent handoff & 31.720 h \\
Spawned agents & 272 \\
Archived agent records & 260 \\
Archive coverage & 95.6\% \\
Input tokens & 771,755,551 \\
Cached input tokens & 744,102,016 \\
Cached input share & 96.4168\% \\
Output tokens & 16,721,326 \\
Total tokens & 788,476,877 \\
Recorded model token cost & US\$10.636892 \\
\bottomrule
\end{tabular}
\end{table}
\endgroup

Archive coverage is the number of archived records divided by the top level spawned agent count. Record level statistics use this archived population.

\subsection{Migration scope and module mapping}
The source comparison maps 13 MESA module directories to the 13 corresponding \texttt{mesa-rs} crates. On the Fortran side, these directories contain \textbf{139,414 physical Fortran lines}, including local \texttt{test/} programs. The mapped Rust crates contain 67,374 library lines and 19,955 crate test lines, giving 87,329 mapped Rust lines in total. Counting all \texttt{.rs} files across the workspace instead gives \textbf{89,946 lines}; the additional 2,617 lines belong to root tests and benchmarks. Both scopes count physical lines, including comments, blank lines and a final line without a terminating newline.

Using these scopes, 139,414 Fortran lines compared with 67,374 Rust library lines gives a physical source size ratio of 2.07$\times$. Including crate-local tests raises the Rust count to 87,329 mapped lines and gives a ratio of 1.60$\times$. Ratios vary widely across modules, partly because the Rust \texttt{eos} and \texttt{kap} crates implement selected functionality from their MESA counterparts. Their source size ratios therefore describe those scoped implementations.

\begin{table}[!htbp]
\centering
\caption{Module level source line counts for MESA and \texttt{mesa-rs}.}\label{tab:migration-modules}
\begin{tabular}{llrrrrr}

\toprule
Layer & Module & Fortran LOC & Rust lib & Rust tests & Rust total & Fortran/Rust \\
\midrule

Foundation & const & 308 & 836 & 90 & 926 & 0.33 \\
Foundation & utils & 3,569 & 1,385 & 370 & 1,755 & 2.03 \\
Foundation & math & 1,020 & 1,474 & 807 & 2,281 & 0.45 \\
Foundation & mtx & 5,206 & 8,244 & 1,728 & 9,972 & 0.52 \\
Foundation & interp\_1d & 5,851 & 2,002 & 937 & 2,939 & 1.99 \\
Foundation & interp\_2d & 16,176 & 5,592 & 1,300 & 6,892 & 2.35 \\
Foundation & num & 18,357 & 10,225 & 3,078 & 13,303 & 1.38 \\
Physics & chem & 4,424 & 3,753 & 900 & 4,653 & 0.95 \\
Physics & rates & 18,547 & 12,304 & 4,975 & 17,279 & 1.07 \\
Physics & neu & 2,263 & 2,222 & 995 & 3,217 & 0.70 \\
Physics & net & 17,799 & 16,332 & 2,429 & 18,761 & 0.95 \\
Physics & eos & 25,817 & 1,426 & 1,514 & 2,940 & 8.78 \\
Physics & kap & 20,077 & 1,579 & 832 & 2,411 & 8.33 \\
\midrule
All & Total & \textbf{139,414} & \textbf{67,374} & \textbf{19,955} & \textbf{87,329} & \textbf{1.60} \\
\bottomrule
\end{tabular}
\end{table}

The final column reports the Fortran/Rust physical line ratio. Testing is described separately through 1,030 static \texttt{\#{}[test]} annotations, 19,955 lines of crate test source and 1,052 passing workspace tests, which measure different aspects of the test infrastructure.

Components outside the migration scope comprise the \texttt{star}, \texttt{astero}, \texttt{binary}, \texttt{adipls}, \texttt{stella} and \texttt{gyre} engines, the full seven source EOS blend, the complete REACLIB dataset, single-precision \texttt{\_sg} variants and \texttt{auto\_diff}. The resulting workspace implements the selected core numerical and physics module chain.

\begin{figure}[H]
\centering
\includegraphics[width=\linewidth]{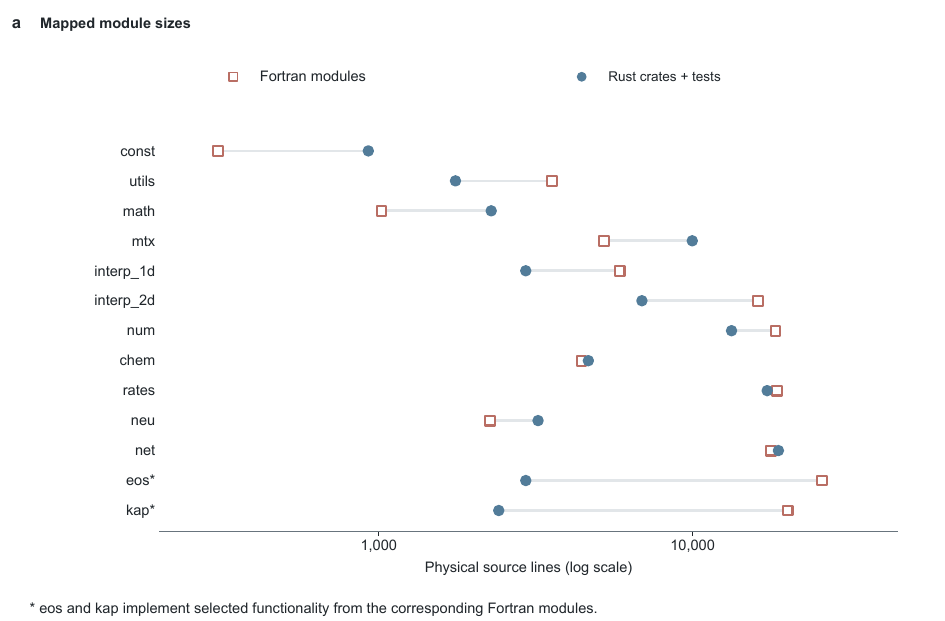}
\caption{\textbf{Module level source line counts for MESA and Rust.} For each of the 13 mapped modules, physical Fortran lines in the module directory, including local test programs, are compared with Rust crate lines, including library code and crate tests. Open red squares denote Fortran modules; filled blue circles denote Rust crates. The logarithmic axis shows both larger and smaller ports. Counts include comments and blank lines; module coverage is specified in the accompanying scope description.}
\label{fig:migration-module-lines}
\end{figure}

\subsection{Run timeline and concurrency}
The archived run lasted 33.22~h, including a root-level handoff at 31.72~h. To examine how work overlapped during this interval, we reconstruct active assignments at one-minute resolution from their episode times. The maximum concurrency is 22 overlapping archived episodes.

\begin{figure}[H]
\centering
\includegraphics[width=\linewidth]{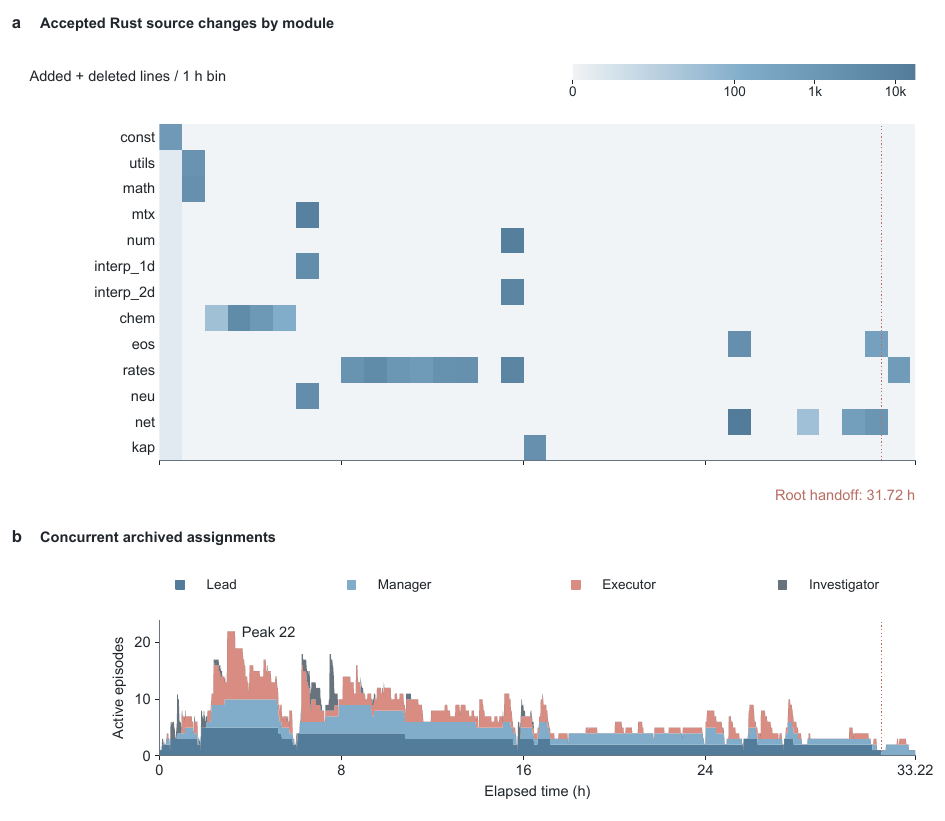}
\caption{\textbf{Accepted code changes and concurrent assignments during migration.} \textbf{a}, Added plus deleted Rust lines in each of the 13 crates, recomputed from first-parent Git diffs at commit \texttt{2ad071e658f8}. One-hour bins start at the root launch; the last bin spans 33 to 33.219 hours. Colour follows $\log(1+\mathrm{added}+\mathrm{deleted})$, with the scale labelled in physical lines. \textbf{b}, Archived assignment concurrency at one-minute resolution, stacked by role; maximum overlap is 22 episodes. Both panels use the same elapsed-time axis. Dotted lines mark the root handoff at 31.72 hours.}
\label{fig:supp-migration-git}\label{fig:supp-migration-concurrency}
\end{figure}

\subsection{Numerical validation and runtime}
The six-workload comparison evaluates selected numerical kernels and pipelines, using the median of 25 runs per implementation after warmup. For the nuclear burn proxy, the timed region integrates a single zone at constant density. Table loading and composition setup occur beforehand, so they lie outside the measured burn integration time.

The timing clocks vary by workload and language. For \texttt{net} and \texttt{kap}, Fortran uses \texttt{system\_clock} and Rust uses \texttt{Instant}; for interpolation, Newton, ROS2 and EOS, Fortran uses \texttt{cpu\_time} and Rust uses \texttt{Instant}. The workloads are reported as single-threaded and CPU-bound, while the implementations use different build toolchains. The resulting runtime ratios therefore characterize the recorded binaries under their stated build configurations and host setup.

Across the batch of 25 runs, every Rust workload had a lower median runtime, with Fortran/Rust ratios ranging from 1.55$\times$ to 6.87$\times$. Numerical comparisons accompany these timings: EOS lookup and Newton workload checksums match exactly, while the other relative checksum differences range from $5.1\times10^{-15}$ to $3.1\times10^{-9}$.

\begingroup
\small
\setlength{\tabcolsep}{3pt}
\begin{table}[!htbp]
\centering
\caption{Numerical agreement and runtime. Speedups use unrounded median runtimes.}\label{tab:migration-performance}
\begin{tabular}{@{}>{\raggedright\arraybackslash}p{0.28\linewidth}rrrr@{}}

\toprule
Workload & Fortran (s) & Rust (s) & Speedup & Checksum difference \\
\midrule

Nuclear burn proxy & 0.446 & 0.287 & 1.55$\times$ & $3.1\times10^{-9}$ \\
EOS lookup & 1.786 & 1.115 & 1.60$\times$ & Identical \\
Opacity lookup & 0.532 & 0.269 & 1.98$\times$ & $1.3\times10^{-13}$ \\
2D interpolation & 0.159 & 0.101 & 1.58$\times$ & $4.9\times10^{-12}$ \\
ROS2 integration & 0.722 & 0.136 & 5.30$\times$ & $5.1\times10^{-15}$ \\
Newton solve & 0.084 & 0.012 & 6.87$\times$ & Identical \\
\bottomrule
\end{tabular}
\end{table}
\endgroup

\Needspace{9\baselineskip}
A separate check repeated the burn proxy 40 times per implementation after three warmup runs, again pinned to CPUs 0 to 3. It yielded medians of 0.2996~s for Fortran and 0.2427~s for Rust, a ratio of 1.23$\times$. Thirty of the 40 Rust timings fell below the Fortran batch median, although individual paired timing vectors are unavailable. Both implementations reported the same integration workload: 8,889 function evaluations, 199 Jacobian evaluations, 199 steps, 198 accepted steps and one rejected step. Their abundance checksums differed by $3.1\times10^{-9}$ relative.

\begingroup
\small
\setlength{\tabcolsep}{4pt}
\begin{table}[!htbp]
\centering
\caption{Separate timing check of 40 runs for the burn proxy. Ratios divide corresponding Fortran and Rust summary statistics.}\label{tab:migration-deep-validation}
\begin{tabular}{@{}>{\raggedright\arraybackslash}p{0.28\linewidth}rrr@{}}

\toprule
Statistic & Fortran (s) & Rust (s) & Fortran/Rust \\
\midrule

Minimum & 0.2897 & 0.2372 & 1.22$\times$ \\
P10 & 0.2922 & 0.2385 & 1.22$\times$ \\
P25 & 0.2927 & 0.2392 & 1.22$\times$ \\
Median & 0.2996 & 0.2427 & 1.23$\times$ \\
Mean & 0.3447 & 0.2735 & 1.26$\times$ \\
P75 & 0.4416 & 0.3609 & 1.22$\times$ \\
P90 & 0.4458 & 0.3729 & 1.20$\times$ \\
Maximum & 0.4490 & 0.3815 & 1.18$\times$ \\
\bottomrule
\end{tabular}
\end{table}
\endgroup

The burn proxy therefore has distinct timing estimates for distinct batches: 1.55$\times$ in the six-workload batch and 1.23$\times$ in the separate batch of 40 runs. The latter archive retains batch summary statistics, but no individual timing vectors for confidence intervals or paired analyses. Additional interleaved measurements yielded 0.82$\times$ under noisy host conditions and approximately 1.22$\times$ in a quieter rerun. These changes document sensitivity to host load and are reported with their respective measurement conditions.

\subsection{Component accuracy and EOS coupling}
\label{sec:mesa-component-checks}
The ROS2 benchmark integrates the Robertson system 50 times from $y(0)=(1,0,0)$ to $t=4\times10^5$, then sums the final $y_1+y_2+y_3$ across runs (Table~\ref{tab:evidence-sources}, \protect\hyperlink{evidence:mesaworkloads2026}{E07}). Mass conservation gives an expected checksum of 50. To assess the components individually, the committed tests evaluate the standard system with rate constants $0.04$, $3\times10^7$ and $10^4$ at $t=10^3$ (Table~\ref{tab:evidence-sources}, \protect\hyperlink{evidence:mesarobertson2026}{E06}), using the reference
\[
y_{\mathrm{ref}}(10^3)=\bigl(0.33687453066069195,\;2.0137023270253555\times10^{-6},\;0.6631234556369591\bigr).
\]
DOP853 established this reference at relative tolerance $10^{-12}$ and absolute tolerance $10^{-14}$, with source-recorded agreement against SciPy LSODA, BDF and Radau. Table~\ref{tab:robertson-components} gives the achieved relative errors recorded alongside the executable tests. These comparisons assess individual components of the Rust solvers against that reference; the six Fortran/Rust workload checksums assess aggregate outputs under their separate protocol.

\begin{table}[!htbp]
\centering\small
\caption{Source-recorded Robertson component errors at $t=10^3$. Both solvers use relative tolerance $10^{-8}$.}\label{tab:robertson-components}
\begin{tabular}{@{}lrrrr@{}}
\toprule
Solver & Absolute tolerance & $y_1$ error & $y_2$ error & $y_3$ error \\
\midrule
DOPRI5 & $10^{-10}$ & $2.5\times10^{-9}$ & $1.2\times10^{-5}$ & $1.2\times10^{-9}$ \\
DOP853 & $10^{-12}$ & $2.9\times10^{-12}$ & $5.1\times10^{-7}$ & $1.0\times10^{-13}$ \\
\bottomrule
\end{tabular}
\end{table}

The high-accuracy run also resolves the transient $y_2$ peak, recording $3.6487146\times10^{-5}$ near $t=4.4555\times10^{-3}$. Executable checks require this peak to fall between $3.6\times10^{-5}$ and $3.7\times10^{-5}$ at a time between $4\times10^{-3}$ and $5\times10^{-3}$. Longer explicit integrations to $10^5$ are marked as ignored tests and run only when explicitly invoked.

An integration test then examines coupling between the EOS table provider and the 22-species nuclear network, comparing it with an ideal-gas-plus-radiation provider at identical burn settings (Table~\ref{tab:evidence-sources}, \protect\hyperlink{evidence:mesaburncoupling2026}{E08}). The initial state uses $\log_{10}T=9.82$ and $\log_{10}\rho=8.2586740078478176$, with integration to $t=0.005$. Because EOS lookup clamps temperature to the table's upper boundary, $\log_{10}T=8.2$, the comparison assesses provider coupling and numerical accounting at that boundary. Both runs enforce abundance conservation within $10^{-6}$, non-negative abundances to $10^{-12}$ and accepted-plus-rejected step accounting.

\begin{table}[!htbp]
\centering\small
\caption{Recorded EOS-provider coupling checks at fixed burn settings.}\label{tab:mesa-eos-coupling}
\begin{tabular}{@{}lrr@{}}
\toprule
Quantity & Ideal gas + radiation & DT table, clamped \\
\midrule
Steps & 352 & 396 \\
Accepted / rejected & 335 / 17 & 377 / 19 \\
RHS evaluations & 14,418 & 16,138 \\
Jacobian evaluations & 352 & 396 \\
Final $\log_{10}T$ & 9.809656 & 9.807399 \\
Mean nuclear energy rate (erg g$^{-1}$ s$^{-1}$) & $-1.246776\times10^{19}$ & $-1.150438\times10^{19}$ \\
\bottomrule
\end{tabular}
\end{table}

The DT run records exactly 16,534 EOS calls, matching the sum of 16,138 RHS evaluations and 396 Jacobian evaluations. Reconstructing the nuclear energy rate from the composition change also agrees within $10^{-6}$ relative. Meanwhile, the two providers yield a maximum absolute abundance difference of $7.453987\times10^{-3}$ and a final $\log_{10}T$ difference of $2.256741\times10^{-3}$. These records show how provider replacement changes the numerical outputs while retaining the checked integration contracts.

\Needspace{30\baselineskip}
\subsection{Golden tests and precision diagnosis}
\label{sec:precision-diagnosis}
The interpolation tests at migration commit \texttt{2ad071e658f8} reconstruct a $31\times35$ grid over $[-\pi,\pi]^2$ and evaluate six coordinates from the Fortran reference driver (Table~\ref{tab:evidence-sources}, \protect\hyperlink{evidence:mesagolden2026}{E05}). The golden output originates from MESA r15140 (commit \texttt{13fe30d5}) using a 2020-era Fortran toolchain. For a value $a$ and reference $b$, the test computes $|a-b|/\max(|a|,|b|)$ after an exact-equality check. Table~\ref{tab:precision-diagnosis} separates the observed discrepancies from the executable acceptance thresholds.

\begin{table}[!htbp]
\centering\small
\caption{Pointwise interpolation checks and recorded precision diagnoses.}\label{tab:precision-diagnosis}
\begin{tabular}{@{}>{\raggedright\arraybackslash}p{.23\linewidth}>{\raggedright\arraybackslash}p{.18\linewidth}>{\raggedright\arraybackslash}p{.14\linewidth}>{\raggedright\arraybackslash}p{.36\linewidth}@{}}
\toprule
Checked quantity & Recorded discrepancy & Assertion & Recorded diagnosis \\
\midrule
12 coordinates; bicubic knot value; four bipm midpoint coefficients & Exact equality & Exact equality & Matching grid construction and coefficient values \\
Three bicubic midpoint derivatives & Maximum $6.4\times10^{-12}$ relative & $10^{-10}$ relative & Current and historical Fortran source agree under the same toolchain; the residual arises against the older golden output \\
18 bicubic values and first derivatives & Maximum $3.05\times10^{-8}$ relative & $10^{-7}$ relative & Implicit \texttt{REAL(4)} Hermite intermediates in the historical Fortran path; double-precision Rust arithmetic \\
Six bipm evaluation values & Five exact; one differs by two ulps ($3.6\times10^{-16}$ relative) & $10^{-13}$ relative & Expression order and intermediate rounding in the monotonic interpolation path \\
\bottomrule
\end{tabular}
\end{table}

Inspection of the compiler intermediate representation confirms the implicit \texttt{REAL(4)} locals in the Fortran path. When the transcription uses explicit double precision, its output matches Rust exactly to the bit; the original mixed-precision path instead reproduces the historical golden value within approximately one ulp. This diagnosis is retained alongside the golden assertions, linking the observed discrepancy to the arithmetic used. Accordingly, the $10^{-7}$ pointwise threshold applies to the historical interpolation comparison. The six workload checksums use the $10^{-8}$ default threshold in the benchmark helper (Table~\ref{tab:evidence-sources}, \protect\hyperlink{evidence:mesahelper}{E21}), under the separate protocol in Table~\ref{tab:migration-performance}.

\subsection{Dependency structure and recursive context}
The mapped dependency graphs contain 142 MESA provider to dependent edges and 41 \texttt{mesa-rs} crate edges. Forty edges appear in both graphs, 102 appear only in the MESA graph and one appears only in Rust. A MESA only edge denotes a provider dependency without a matching direct Rust crate edge in the mapped graph.

\begingroup
\small
\setlength{\tabcolsep}{3pt}
\begin{table}[!htbp]
\centering
\caption{Mapped dependency counts for MESA and \texttt{mesa-rs}.}\label{tab:migration-dependencies}
\begin{tabular}{@{}>{\raggedright\arraybackslash}p{0.20\linewidth}r >{\raggedright\arraybackslash}p{0.66\linewidth}@{}}

\toprule
Category & Edges & Interpretation \\
\midrule

MESA & 142 & Provider to dependent edges parsed from \texttt{INTERNAL\_DEPENDS\_ON} \\
\texttt{mesa-rs} & 41 & Mapped crate dependencies parsed from Cargo.toml \\
Retained & 40 & Edges present in both mapped graphs \\
MESA only & 102 & No matching direct Rust crate dependency in the mapped graph \\
Rust only & 1 & Dependency introduced in \texttt{mesa-rs} \\
\bottomrule
\end{tabular}
\end{table}
\endgroup

\begin{figure}[H]
\centering
\includegraphics[width=\linewidth]{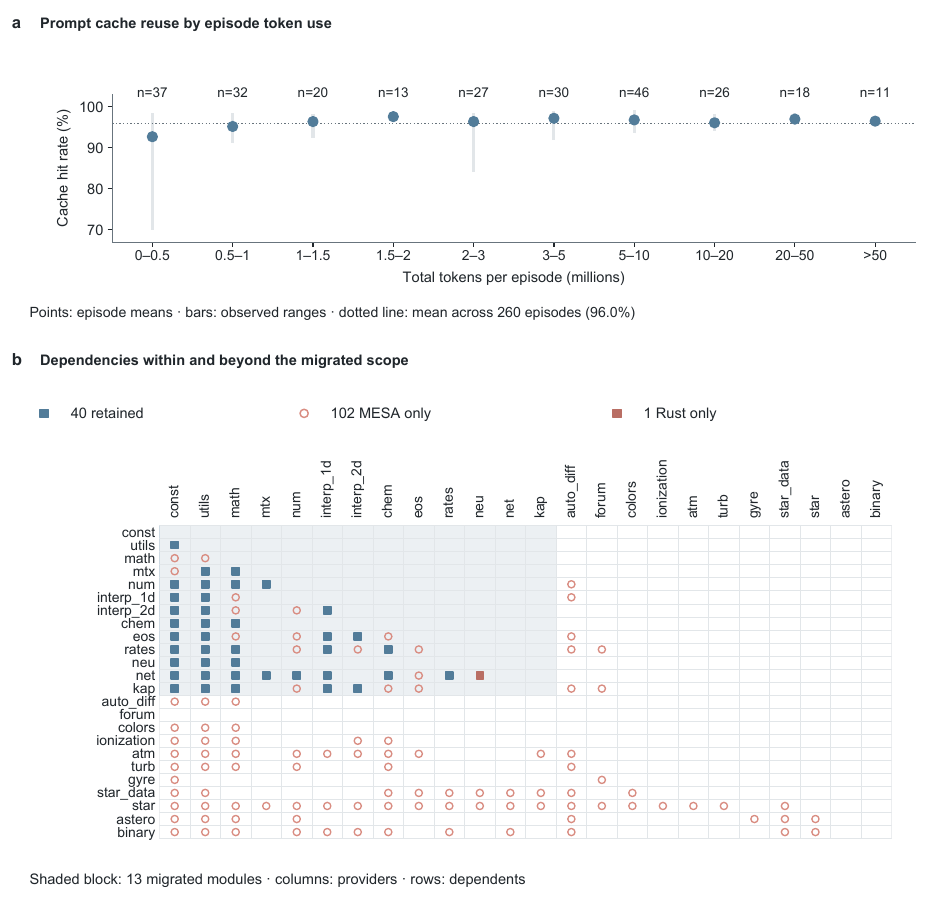}
\caption{\textbf{Prompt cache reuse and mapped module dependencies.} \textbf{a}, Points show the arithmetic mean of episode cache hit rates in each total-token group; vertical bars span observed minima and maxima, and labels give sample sizes. The dotted line is the mean across 260 archived episodes (96.0\%). Table~\ref{tab:migration-resources} separately reports cached input divided by total input at run level (96.4168\%). \textbf{b}, Columns identify providers and rows identify dependents. The shaded block contains the 13 migrated modules; surrounding rows and columns retain the wider MESA dependency graph. Symbols distinguish 40 shared edges, 102 MESA-only edges and one Rust-only edge.}
\label{fig:supp-mesa-structure}
\end{figure}

Archived assignments span 34 observed context paths. Adding the necessary parent paths yields a hierarchy of 37 nodes, and twenty-six observed paths contain \texttt{CONTEXT.md}. Their Git history records 26 creations and 62 later accepted updates affecting 19 files, showing how shared context changed during redevelopment. These measurements cover accepted edits; file reads and the performance effects of individual updates were not measured.

\begin{figure}[H]
\centering
\includegraphics[width=\linewidth]{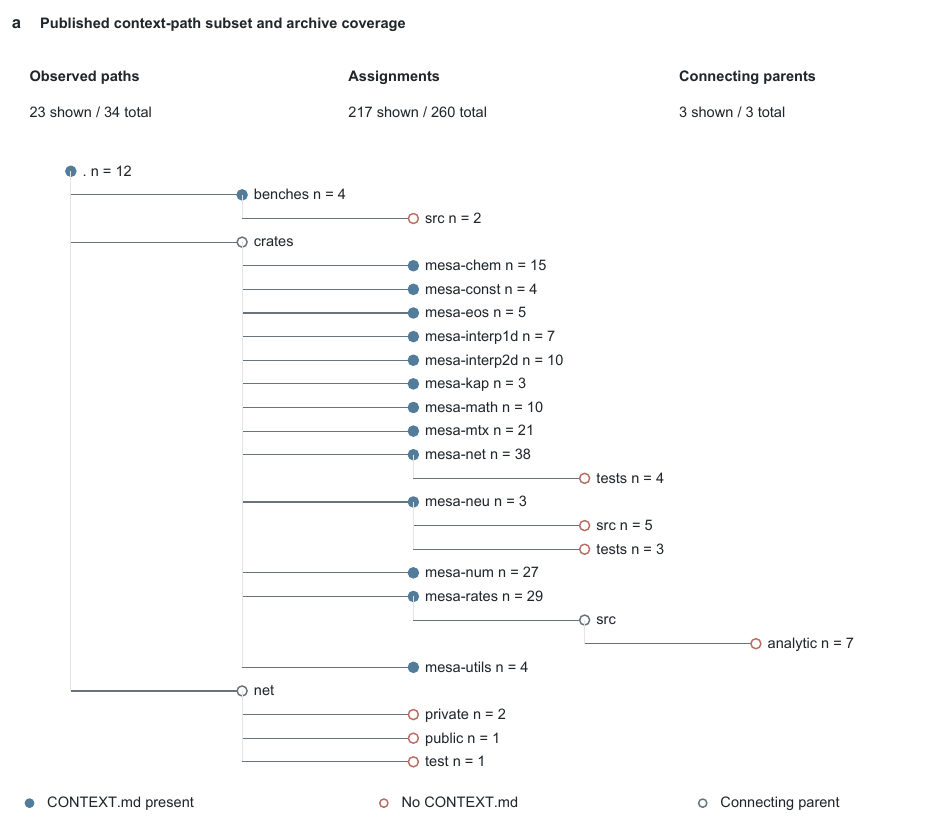}
\caption{\textbf{Context-path subset and archive coverage.} The displayed hierarchy contains 23 observed paths covering 217 archived assignments, plus three connecting parents. It reproduces the path subset in the original experiment figure. The complete archive contains 34 observed paths and 260 assignments; the remaining 11 paths account for 43 assignments. Filled markers denote paths with \texttt{CONTEXT.md}, open red markers paths without it, and open grey markers connecting parents. Labels give assignment counts at each displayed path.}
\label{fig:supp-migration-scope}
\end{figure}

\begin{figure}[H]
\centering
\includegraphics[width=\linewidth]{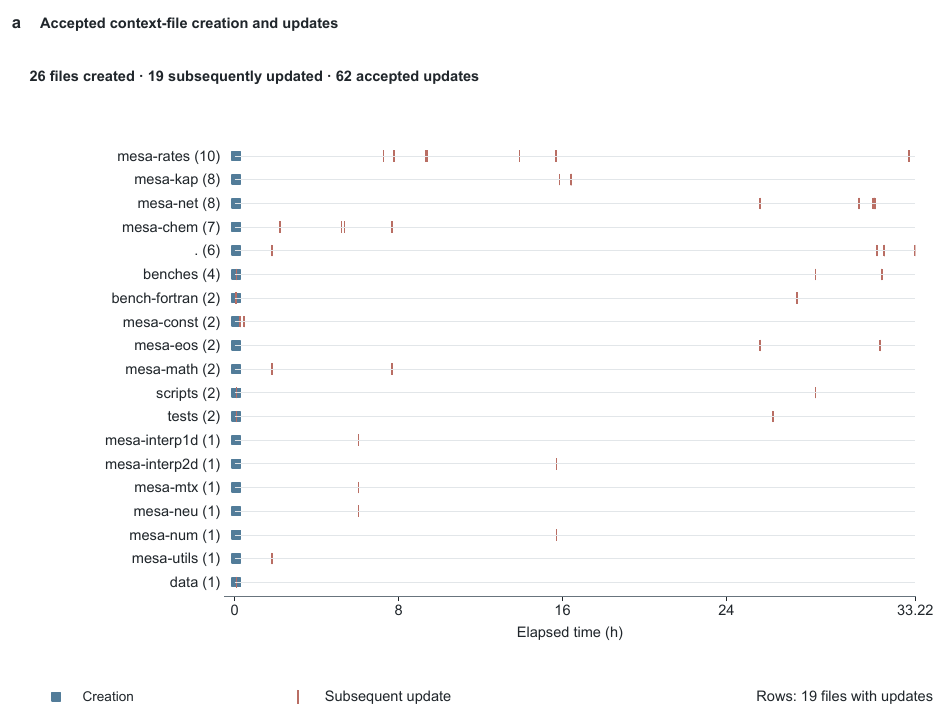}
\caption{\textbf{Accepted changes to shared context files.} First-parent Git history records 26 file creations and 62 subsequent updates. The plot shows the 19 files with subsequent updates; parenthesized counts give their update totals. Blue squares mark creation and red ticks mark subsequent updates. The other seven files were created without later accepted updates during this run.}
\label{fig:supp-migration-context}
\end{figure}

The run produced 13 Rust crates for the reported MESA core module scope. The final workspace passed 1,052 tests, and the tested numerical workloads showed the agreement reported above.

\section{Formation: a software WebGL renderer}
\label{sec:webgl}
\subsection{Version, implementation and measurement scope}
The WebGL verification record is pinned to repository revision \texttt{67f620392b66} and tests the committed \texttt{renderer.js} bundle from \texttt{c4ffbff} (Table~\ref{tab:evidence-sources}, \protect\hyperlink{evidence:webglfinal2026}{E09}). The renderer is written in TypeScript and distributed as an ES2019 JavaScript bundle with zero runtime dependencies. Within that implementation, source contracts separate shader compilation, rasterization, API state and presentation, specifying data layouts and error behaviour at their interfaces (Table~\ref{tab:evidence-sources}, \protect\hyperlink{evidence:webglarchitecture2026}{E10}).

The project evidence record reports DeepSeek V4 Flash and US\$36 in model token charges (Table~\ref{tab:evidence-sources}, \protect\hyperlink{evidence:webglevidence2026}{E11}). The 65.9 hour development interval is the difference between the first and last agent commit timestamps, \texttt{2026-08-26T16:58:38Z} and \texttt{2026-08-29T10:55:17Z}. Episode counts and cumulative agent hours are unreported. The conformance runs lasted 1,420.5 seconds for WebGL1 and 526.6 seconds for WebGL2, using four browser workers. The harness used a 60 second idle timeout, page-specific slow-test overrides and a 35 minute wall-clock cap.

\subsection{Conformance manifests and repair records}
The CTS tree identifies version 2.0.1 (beta). Its \texttt{00\_test\_list.txt} manifests in \texttt{conformance/} and \texttt{conformance2/} contain 835 and 1,184 graded pages, respectively. Since each page aggregates its own subtests, Table~\ref{tab:webgl-conformance} reports page-level outcomes. All 2,019 graded pages record \texttt{rendererActive: true}, and neither suite reports skipped subtests.

\begin{table}[!htbp]
\centering\small
\caption{Recorded WebGL conformance outcomes for bundle \texttt{c4ffbff}.}\label{tab:webgl-conformance}
\begin{tabular}{@{}lrrrrr@{}}
\toprule
Manifest & Passed & Failed & Timeout & Error & Pages \\
\midrule
WebGL1 & 833 & 2 & 0 & 0 & 835 \\
WebGL2 & 1,184 & 0 & 0 & 0 & 1,184 \\
Combined & 2,017 & 2 & 0 & 0 & 2,019 \\
\bottomrule
\end{tabular}
\end{table}

{\small
\begin{table}[!htbp]
\centering
\caption{Two WebGL2 repairs, with subtest counts before and after the fixes.}\label{tab:webgl-repairs}
\begin{tabular}{@{}>{\raggedright\arraybackslash}p{.20\linewidth}>{\raggedright\arraybackslash}p{.34\linewidth}>{\raggedright\arraybackslash}p{.17\linewidth}>{\raggedright\arraybackslash}p{.20\linewidth}@{}}

\toprule
Page / behaviour & Accepted change & Subtests before $\rightarrow$ after & Repair commits \\
\midrule
Origin-clean OffscreenCanvas image upload & Check origin-clean status before texture binding at all six image-source entry points & 9 pass, 4 fail $\rightarrow$ 13 pass, 0 fail & \texttt{7d8718a} \\
High-bit-depth texture input & Preserve 16-bit PNG samples through decoding; correct RGBA channel handling & 2 pass, 1 fail $\rightarrow$ 3 pass, 0 fail & \texttt{89e4559}, \texttt{9bd7811} \\
\bottomrule
\end{tabular}
\end{table}}

The preceding bundle \texttt{e3565d4} passed 2,015/2,019 pages, comprising 833/835 WebGL1 and 1,182/1,184 WebGL2 pages. The final full run subsequently recovered both WebGL2 failures while preserving the prior passes. The two nonpassing WebGL1 pages remained: low-alpha colour readback had 81 passing and four failing subtests, while double-underscore shader identifiers had one passing and two failing subtests. Native WebGL controls recorded matching failures (Table~\ref{tab:evidence-sources}, \protect\hyperlink{evidence:webglfinal2026}{E09}).

Visual comparisons complement conformance testing by examining image agreement on selected scenes. The project evidence record reports 16/38 selected three.js pages passing a strict 0.1\% pixel threshold and 13/16 selected Babylon.js scenes (Table~\ref{tab:evidence-sources}, \protect\hyperlink{evidence:webglevidence2026}{E11}). These scene outcomes retain their own denominators alongside the CTS page outcomes in Table~\ref{tab:webgl-conformance}.

\subsection{Conformance progression and scene-level repairs}
\label{sec:webgl-visual-records}
Three committed full-suite reports follow the same 835-page WebGL1 and 1,184-page WebGL2 manifests, allowing page outcomes to be compared across revisions (Table~\ref{tab:evidence-sources}, \protect\hyperlink{evidence:webglstages2026}{E15}, \protect\hyperlink{evidence:webglfinal2026}{E09}). Table~\ref{tab:webgl-progression} records that progression. All six suite runs report zero timeouts, zero errors and active software rendering on every page. Between \texttt{ce8a79c} and \texttt{b56a985}, WebGL1 recovers eight pages, with all remaining failures drawn from the earlier failure set. WebGL2, however, recovers 39 pages while introducing five new failures, for a net gain of 34 passes. The final bundle subsequently passes every WebGL2 page.

\begin{table}[!htbp]
\centering\small
\caption{Recorded CTS progression on fixed page manifests.}\label{tab:webgl-progression}
\begin{tabular}{@{}lrrr@{}}
\toprule
Bundle & WebGL1 passed & WebGL2 passed & Combined passed \\
\midrule
\texttt{ce8a79c} & 824/835 & 1,130/1,184 & 1,954/2,019 \\
\texttt{b56a985} & 832/835 & 1,164/1,184 & 1,996/2,019 \\
\texttt{c4ffbff} & 833/835 & 1,184/1,184 & 2,017/2,019 \\
\bottomrule
\end{tabular}
\end{table}

Babylon.js comparisons use pixelmatch with a pixel-level threshold of 0.035 and scene-specific limits on the differing-pixel ratio (Table~\ref{tab:evidence-sources}, \protect\hyperlink{evidence:webglbabylon2026}{E13}). To compare results against those limits, Fig.~\ref{fig:webgl-development}b divides each measured ratio by its scene budget; Table~\ref{tab:babylon-repair-pairs} preserves the original ratios. The associated repairs address pre-blend clamping (\texttt{a18b9f8}), matrix aliasing (\texttt{fdeebd1}) and the Euclidean-norm level-of-detail footprint (\texttt{ca40477}). After these repairs, the curated suite passes 13/16 scenes. Hill Valley (3.08\%), Heart (1.80\%) and Windows cafe (1.35\%) remain above their 1.1\% budgets. These measurements use the completed asset cache, while asset-load failures are recorded separately.

\begin{table}[!htbp]
\centering\small
\caption{Recorded Babylon.js scene errors before and after the repair family. Values are percentages of differing pixels.}\label{tab:babylon-repair-pairs}
\begin{tabular}{@{}lrrrl@{}}
\toprule
Scene & Before & After & Budget & Final outcome \\
\midrule
The car & 7.89 & 2.78 & 5.00 & Pass \\
Sponza & 7.64 & 0.93 & 1.10 & Pass \\
Hill Valley & 7.54 & 3.08 & 1.10 & Fail \\
Flat2009 & 2.40 & 0.76 & 1.10 & Pass \\
\bottomrule
\end{tabular}
\end{table}

The historical three.js evaluation uses the recorded v0.185-dev example tree and passes 16/38 selected pages (Table~\ref{tab:evidence-sources}, \protect\hyperlink{evidence:webglthreejs2026}{E14}). A page passes when fewer than 0.1\% of compared pixels exceed a normalized squared RGB distance of 0.01. The driver downsamples the captured $800\times500$ frame to $400\times250$ for comparison. Commit \texttt{6b068f7} normalizes JPEG goldens to RGBA before comparison; the values in Table~\ref{tab:threejs-scene-errors} use this corrected metric. The source gives explicit final percentages for 35 pages. The other three pages remain in the 38-page suite and have separately recorded golden/native controls: \path{multisampled_renderbuffers}, \path{postprocessing_fxaa} and \texttt{lines\_fat}.

{\small
\setlength{\tabcolsep}{3pt}
\begin{table}[!htbp]
\centering
\caption{Explicit three.js scene errors in the committed final record. The common \texttt{webgl\_} prefix is omitted; percentages retain the source's three-decimal precision.}\label{tab:threejs-scene-errors}
\begin{tabular}{@{}>{\raggedright\arraybackslash}p{.42\linewidth}r>{\raggedright\arraybackslash}p{.32\linewidth}@{}}

\toprule
Scene & Differing pixels (\%) & Result at 0.1\% threshold \\
\midrule
geometry\_terrain & 0.000 & Pass \\
materials\_video & 0.000 & Pass \\
morphtargets & 0.000 & Pass \\
postprocessing\_afterimage & 0.000 & Pass \\
postprocessing\_unreal\_bloom & 0.000 & Pass \\
shadowmap\_vsm & 0.000 & Pass \\
materials\_displacementmap & 0.001 & Pass \\
morphtargets\_horse & 0.001 & Pass \\
geometry\_text & 0.003 & Pass \\
materials\_normalmap & 0.011 & Pass \\
buffergeometry\_instancing & 0.020 & Pass \\
materials\_cubemap\_dynamic & 0.024 & Pass \\
postprocessing\_dof & 0.044 & Pass \\
materials\_envmaps & 0.052 & Pass \\
points\_waves & 0.062 & Pass \\
points\_dynamic & 0.096 & Pass \\
tonemapping & 0.129 & Fail \\
animation\_skinning\_blending & 0.131 & Fail \\
mirror & 0.131 & Fail \\
points\_billboards & 0.176 & Fail \\
lights\_hemisphere & 0.202 & Fail \\
materials\_physical\_transmission & 0.213 & Fail \\
instancing\_raycast & 0.235 & Fail \\
instancing\_dynamic & 0.271 & Fail \\
postprocessing\_ssao & 0.577 & Fail \\
materials\_texture\_anisotropy & 0.781 & Fail \\
animation\_skinning\_morph & 0.877 & Fail \\
shadowmap & 0.892 & Fail \\
geometry\_spline\_editor & 1.180 & Fail \\
animation\_keyframes & 1.590 & Fail \\
interactive\_raycasting\_points & 2.060 & Fail \\
buffergeometry & 2.314 & Fail \\
points\_sprites & 2.458 & Fail \\
materials\_physical\_clearcoat & 4.763 & Fail \\
geometry\_shapes & 4.784 & Fail \\
\bottomrule
\end{tabular}
\end{table}}

Native controls help distinguish rendering differences from differences involving the golden images or their environments. For \path{instancing_dynamic} and \path{instancing_raycast}, native output gives 0.031\% and 0.001\% differing pixels, compared with 0.271\% and 0.235\% for the software renderer. In \path{postprocessing_fxaa}, software and native output differ by 0.053\%, while both differ from the golden by approximately 22.4\%. For \path{multisampled_renderbuffers}, native versus golden differs by 60.79\% and software versus native by 10.13\%. Finally, software and native output for \texttt{lines\_fat} differ from the golden by 12.15\% and 12.55\%, respectively. Each result retains its particular comparison and denominator.

\FloatBarrier
\subsection{Fixed-version rendering captures}\label{sec:webgl-captures}
The captures in Fig.~\ref{fig:webgl-development}a provide a fixed-version comparison separate from the archived visual-suite evaluation. They were taken on 16 November 2026 using \path{renderer.js} from commit \texttt{67f620392b66}, which contains the tested \texttt{c4ffbff} bundle (Table~\ref{tab:evidence-sources}, \protect\hyperlink{evidence:webglbundle2026}{E12}). The renderer file's SHA-256 is \path{7c3a185e22f27f9dc888f5fc0bd065c6d92dde87ddb954de108483e01e2c1684}. The two scenes, \path{webgl_geometry_shapes} and \path{webgl_instancing_dynamic}, come from three.js r184 (Table~\ref{tab:evidence-sources}, \protect\hyperlink{evidence:threeexamples2026}{E20}), so the captures retain their own scene version and comparison protocol.

Both renderers ran in Chromium 153.0.8010.12 with an $800\times500$ viewport and device scale factor one. The reference used Chromium's native WebGL API with ANGLE SwiftShader. The software branch intercepted WebGL context creation with the committed factory, and each capture recorded its successful activation. Animation time was fixed at zero; two requested animation frames were executed after assets loaded. Scene-level random calls used a dedicated deterministic stream initialized to $\pi/4$, with each value given by the fractional part of $10^4\sin(s)$ followed by $s\leftarrow s+1$. Camera and projection matrices, geometry positions, object transforms, instance matrices and instance colours matched between each reference/software pair by SHA-256. Captures recorded no page or console errors.

The local views in Fig.~\ref{fig:webgl-development}a use matched $256\times160$ pixel crops: coordinates $(330,210)$ to $(586,370)$ for procedural geometry and $(270,285)$ to $(526,445)$ for dynamic instancing, measured from the top left of each $800\times500$ canvas. Reference, software and difference crops use the same coordinates. The numerical summaries, however, cover all 400,000 full-frame pixels rather than just those crops. The source data retain the empirical cumulative distribution at each threshold $j/255$, with mean differences of 0.013499 for geometry and 0.098008 for instancing. Each summary describes one fixed frame of its scene.

The PNGs retain the complete canvas without image enhancement, with HTML labels and performance overlays excluded. Difference images then visualize the maximum absolute RGB channel difference divided by 255 on a linear scale from zero to one, shared by both scenes. The capture manifest records file hashes, browser version, download URLs and capture parameters, and is archived with the script and original PNGs.

\section{EvoX Compiler: benchmarks and verification}
\label{sec:evoxc}
\subsection{Version and release accounting}
The EvoX Compiler case in main-text Section~\ref{sec:evoxc-main} and Fig.~\ref{fig:evoxc-results} uses v0.1.0 at commit \texttt{2b289f034f6f}. Its release README reports results for the colocated \texttt{tests/} and \texttt{evoxc\_hf/tests/} suites under two dependency configurations (Table~\ref{tab:evidence-sources}, \protect\hyperlink{evidence:evoxc2026}{E16}). The latter suite belongs to a separately distributed HuggingFace integration package that uses the compiler's public API. Table~\ref{tab:evoxc-tests} therefore retains the scope of each configuration together with its passed, skipped, expected-failure and failed counts.

\begin{table}[!htbp]
\centering\small
\caption{EvoX Compiler v0.1.0 reported test outcomes.}\label{tab:evoxc-tests}
\begin{tabular}{@{}lrrrr@{}}
\toprule
Configuration & Passed & Skipped & Expected failures & Failed \\
\midrule
All extras (IREE, TVM, torch) & 7,511 & 77 & 4 & 0 \\
Compiler-free & 7,086 & 129 & 4 & 0 \\
\bottomrule
\end{tabular}
\end{table}

\Needspace{10\baselineskip}
\subsection{Program-merging benchmark records}
Figure~\ref{fig:evoxc-results} draws on the performance summaries retained in the v0.1.0 repository (Table~\ref{tab:evidence-sources}, \protect\hyperlink{evidence:evoxcperf2026}{E19}). For the 32-program workload, 10-deep elementwise chains share a length-64 vector and are merged with \texttt{switch\_ops=True}. The record reports best-of-five execution times, with outputs bit-identical to separate execution. Merging reduces the operation count from 544 to 133; the compilation comparison correspondingly measures 32 separate IREE CPU compiles against one merged compile.

The second workload contains 100 divergent expression programs with transcendental operations and two length-64 float32 inputs. It compares separate execution with a flat merge and automatic banding (\texttt{bands="auto"}). Here, the archived summary specifies best-of-$N$ timing, but does not record the repetition count, individual timings or CPU model. Figure results therefore retain the reported point values and keep this workload separate from the first, with distinct linear axes for NumPy and IREE CPU execution times.

\Needspace{10\baselineskip}
\subsection{Graph test inputs and assertions}
The \texttt{batch\_functions} integration test checks how graph transformation preserves the association between functions and their outputs. It uses $x+1$, $2x$ and $x-3$ with a static function axis and relative and absolute tolerances of $10^{-5}$. Compiled output is compared with both the NumPy evaluation of the merged graph and stacked NumPy evaluations of the original programs. Further assertions require StableHLO export to contain the selection operation for divergent programs and check that batching identical programs agrees with \texttt{vmap} (Table~\ref{tab:evidence-sources}, \protect\hyperlink{evidence:evoxcbatch2026}{E17}). Together, these checks specify the row-to-function mapping through transformation, lowering and execution.

\subsection{Lifecycle errors and test doubles}
Backend lifecycle tests use a test-only PJRT plugin that accumulates completion-event handles across successive runs. Synchronization must await and destroy every handle, clear the pending list and remain idempotent when called again. Error-path assertions additionally check cleanup after await or destroy failures and require an explicit error if the plugin supplies no completion event (Table~\ref{tab:evidence-sources}, \protect\hyperlink{evidence:evoxcsync2026}{E18}). Because this plugin returns zero-filled outputs, these tests assess lifecycle and error semantics; the graph tests above supply the numerical checks.

\clearpage
\section{Versioned artifacts and reproduction entry points}
\label{sec:reproduction}
The repository snapshots in Table~\ref{tab:repository-commits} identify the software versions used in each study. Table~\ref{tab:evidence-sources} maps the evidence identifiers cited throughout the paper to the corresponding test records, source files and versioned reports. File links resolve to the recorded commit or release tag; the project evidence page retains its access date.

\Needspace{7\baselineskip}
With the recorded CTS tree and harness dependencies installed, WebGL verification runs from the renderer repository root as follows:
\begin{quote}\small\ttfamily
npx tsx tests/conformance/run.ts\newline
\hspace*{1em}--renderer ./renderer.js --suite webgl1\newline
\hspace*{1em}--report tests/reports/webgl1-FINAL2.json
\end{quote}
The WebGL2 command substitutes \texttt{webgl2} for the suite and report basename. The committed report preserves the aggregate counts and repair identifiers; detailed JSON outputs are gitignored by the source project.

Genesis v0.14.1, released on 8 November 2026, is the implementation release snapshot for this report~\cite{genesisrelease2026}. Its changes include canonical provider/model selection, serialization of agent archive settings, consolidation of consecutive context-file edits, and a three-turn recovery window at the agent turn limit. Historical measurements remain attached to their recorded run configurations and repository snapshots.

\clearpage
\subsection{Evidence source index}
\begingroup\small
\setlength{\tabcolsep}{5pt}
\begin{longtable}{@{}>{\raggedright\arraybackslash}p{.07\linewidth}>{\raggedright\arraybackslash}p{.45\linewidth}>{\raggedright\arraybackslash}p{.42\linewidth}@{}}
\caption{Versioned sources for experimental evidence. Each identifier links to its source row from the main text and Supplementary Information.}\label{tab:evidence-sources}\\
\toprule
ID & Evidence and project & Version and linked source \\
\midrule\endfirsthead
\multicolumn{3}{@{}l}{\textbf{Table \thetable.} Versioned sources for experimental evidence (continued).}\\[5pt]
\toprule ID & Evidence and project & Version and linked source \\
\midrule\endhead
\midrule\multicolumn{3}{r@{}}{Continued on next page}\\\endfoot
\bottomrule\endlastfoot
\hypertarget{evidence:jcccsmith2026}{E01} & Csmith differential testing: repair and revalidation records.\newline\cite{jccrepo2026} & \texttt{a8c116ec7ed3}\hfill\href{https://github.com/EMI-Group/genesis-demo-jcc/blob/a8c116ec7ed3d59479ad3be8fed3e63384f50a87/tools/csmith/CONTEXT.md}{\sffamily GitHub source}\newline\href{https://github.com/EMI-Group/genesis-demo-jcc/blob/a8c116ec7ed3d59479ad3be8fed3e63384f50a87/tools/csmith/CONTEXT.md}{\nolinkurl{tools/csmith/CONTEXT.md}} \\[7pt]
\hypertarget{evidence:jccllvmaudit2026}{E02} & LLVM SingleSource run-failure categorization.\newline\cite{jccrepo2026} & \texttt{10077f656861}\hfill\href{https://github.com/EMI-Group/genesis-demo-jcc/blob/10077f65686102deef2c228e8a49b0569218ec48/tests/llvm_run_failure_categories.md}{\sffamily GitHub source}\newline\href{https://github.com/EMI-Group/genesis-demo-jcc/blob/10077f65686102deef2c228e8a49b0569218ec48/tests/llvm_run_failure_categories.md}{\nolinkurl{tests/llvm_run_failure_categories.md}} \\[7pt]
\hypertarget{evidence:jccdsprotocol2026}{E03} & DeepSeek continuation: test runner and collection history.\newline\cite{jccrepo2026} & \texttt{10077f656861}\hfill\href{https://github.com/EMI-Group/genesis-demo-jcc/blob/10077f65686102deef2c228e8a49b0569218ec48/tests/CONTEXT.md}{\sffamily GitHub source}\newline\href{https://github.com/EMI-Group/genesis-demo-jcc/blob/10077f65686102deef2c228e8a49b0569218ec48/tests/CONTEXT.md}{\nolinkurl{tests/CONTEXT.md}} \\[7pt]
\hypertarget{evidence:jccglmprotocol2026}{E04} & GLM continuation: final LLVM SingleSource runner.\newline\cite{jccrepo2026} & \texttt{ab58a91460ad}\hfill\href{https://github.com/EMI-Group/genesis-demo-jcc/blob/ab58a91460adb24e9349a554e02d4522d3ae2db4/tests/run_llvm_testsuite.py}{\sffamily GitHub source}\newline\href{https://github.com/EMI-Group/genesis-demo-jcc/blob/ab58a91460adb24e9349a554e02d4522d3ae2db4/tests/run_llvm_testsuite.py}{\nolinkurl{tests/run_llvm_testsuite.py}} \\[7pt]
\hypertarget{evidence:mesagolden2026}{E05} & Interpolation golden tests and precision diagnoses.\newline\cite{mesarepo2026} & \texttt{2ad071e658f8}\hfill\href{https://github.com/EMI-Group/genesis-demo-mesa-rs/blob/2ad071e658f8d4de21aa9f0e906592c364e10023/crates/mesa-interp2d/tests/golden.rs}{\sffamily GitHub source}\newline\href{https://github.com/EMI-Group/genesis-demo-mesa-rs/blob/2ad071e658f8d4de21aa9f0e906592c364e10023/crates/mesa-interp2d/tests/golden.rs}{\nolinkurl{crates/mesa-interp2d/tests/golden.rs}} \\[7pt]
\hypertarget{evidence:mesarobertson2026}{E06} & Robertson component tests and numerical references.\newline\cite{mesarepo2026} & \texttt{2ad071e658f8}\hfill\href{https://github.com/EMI-Group/genesis-demo-mesa-rs/blob/2ad071e658f8d4de21aa9f0e906592c364e10023/crates/mesa-num/tests/test_robertson.rs}{\sffamily GitHub source}\newline\href{https://github.com/EMI-Group/genesis-demo-mesa-rs/blob/2ad071e658f8d4de21aa9f0e906592c364e10023/crates/mesa-num/tests/test_robertson.rs}{\nolinkurl{crates/mesa-num/tests/test_robertson.rs}} \\[7pt]
\hypertarget{evidence:mesaworkloads2026}{E07} & Fortran and Rust benchmark workload contracts.\newline\cite{mesarepo2026} & \texttt{2ad071e658f8}\hfill\href{https://github.com/EMI-Group/genesis-demo-mesa-rs/blob/2ad071e658f8d4de21aa9f0e906592c364e10023/bench-fortran/WORKLOADS.md}{\sffamily GitHub source}\newline\href{https://github.com/EMI-Group/genesis-demo-mesa-rs/blob/2ad071e658f8d4de21aa9f0e906592c364e10023/bench-fortran/WORKLOADS.md}{\nolinkurl{bench-fortran/WORKLOADS.md}} \\[7pt]
\hypertarget{evidence:mesaburncoupling2026}{E08} & EOS-provider coupling and nuclear burn integration tests.\newline\cite{mesarepo2026} & \texttt{2ad071e658f8}\hfill\href{https://github.com/EMI-Group/genesis-demo-mesa-rs/blob/2ad071e658f8d4de21aa9f0e906592c364e10023/tests/one_zone_burn.rs}{\sffamily GitHub source}\newline\href{https://github.com/EMI-Group/genesis-demo-mesa-rs/blob/2ad071e658f8d4de21aa9f0e906592c364e10023/tests/one_zone_burn.rs}{\nolinkurl{tests/one_zone_burn.rs}} \\[7pt]
\hypertarget{evidence:webglfinal2026}{E09} & WebGL CTS final verification record.\newline\cite{webglrepo2026} & \texttt{67f620392b66}\hfill\href{https://github.com/BillHuang2001/tbench-wasm/blob/67f620392b667714ba028f320d0e24f8ed32a16a/tests/conformance/FINAL.md}{\sffamily GitHub source}\newline\href{https://github.com/BillHuang2001/tbench-wasm/blob/67f620392b667714ba028f320d0e24f8ed32a16a/tests/conformance/FINAL.md}{\nolinkurl{tests/conformance/FINAL.md}} \\[7pt]
\hypertarget{evidence:webglarchitecture2026}{E10} & Software WebGL renderer: source architecture and interface contracts.\newline\cite{webglrepo2026} & \texttt{67f620392b66}\hfill\href{https://github.com/BillHuang2001/tbench-wasm/blob/67f620392b667714ba028f320d0e24f8ed32a16a/src/CONTEXT.md}{\sffamily GitHub source}\newline\href{https://github.com/BillHuang2001/tbench-wasm/blob/67f620392b667714ba028f320d0e24f8ed32a16a/src/CONTEXT.md}{\nolinkurl{src/CONTEXT.md}} \\[7pt]
\hypertarget{evidence:webglevidence2026}{E11} & Software WebGL renderer: development and verification evidence.\newline EvoX Genesis contributors & \href{https://genesis.evox.group/evidence/tbench-wasm/}{Project evidence page}\newline Accessed 2026-11-13 \\[7pt]
\hypertarget{evidence:webglbundle2026}{E12} & Committed software WebGL renderer bundle.\newline\cite{webglrepo2026} & \texttt{67f620392b66}\hfill\href{https://github.com/BillHuang2001/tbench-wasm/blob/67f620392b667714ba028f320d0e24f8ed32a16a/renderer.js}{\sffamily GitHub source}\newline\href{https://github.com/BillHuang2001/tbench-wasm/blob/67f620392b667714ba028f320d0e24f8ed32a16a/renderer.js}{\nolinkurl{renderer.js}} \\[7pt]
\hypertarget{evidence:webglbabylon2026}{E13} & Babylon.js visual comparisons and scene repair records.\newline\cite{webglrepo2026} & \texttt{67f620392b66}\hfill\href{https://github.com/BillHuang2001/tbench-wasm/blob/67f620392b667714ba028f320d0e24f8ed32a16a/tests/babylon/CONTEXT.md}{\sffamily GitHub source}\newline\href{https://github.com/BillHuang2001/tbench-wasm/blob/67f620392b667714ba028f320d0e24f8ed32a16a/tests/babylon/CONTEXT.md}{\nolinkurl{tests/babylon/CONTEXT.md}} \\[7pt]
\hypertarget{evidence:webglthreejs2026}{E14} & three.js final visual comparisons and native controls.\newline\cite{webglrepo2026} & \texttt{67f620392b66}\hfill\href{https://github.com/BillHuang2001/tbench-wasm/blob/67f620392b667714ba028f320d0e24f8ed32a16a/tests/threejs/CONTEXT.md}{\sffamily GitHub source}\newline\href{https://github.com/BillHuang2001/tbench-wasm/blob/67f620392b667714ba028f320d0e24f8ed32a16a/tests/threejs/CONTEXT.md}{\nolinkurl{tests/threejs/CONTEXT.md}} \\[7pt]
\hypertarget{evidence:webglstages2026}{E15} & WebGL conformance: versioned full-suite baseline reports.\newline\cite{webglrepo2026} & \texttt{67f620392b66}\hfill\href{https://github.com/BillHuang2001/tbench-wasm/tree/67f620392b667714ba028f320d0e24f8ed32a16a/tests/conformance}{\sffamily GitHub source}\newline\href{https://github.com/BillHuang2001/tbench-wasm/tree/67f620392b667714ba028f320d0e24f8ed32a16a/tests/conformance}{\nolinkurl{tests/conformance}} \\[7pt]
\hypertarget{evidence:evoxc2026}{E16} & EvoX Compiler v0.1.0: architecture and release test summary.\newline\cite{evoxc2026} & \texttt{2b289f034f6f}\hfill\href{https://github.com/EMI-Group/evoxc/blob/2b289f034f6f8e16b6c9b55932dfed76cc54690e/README.md}{\sffamily GitHub source}\newline\href{https://github.com/EMI-Group/evoxc/blob/2b289f034f6f8e16b6c9b55932dfed76cc54690e/README.md}{\nolinkurl{README.md}} \\[7pt]
\hypertarget{evidence:evoxcbatch2026}{E17} & Graph batching: backend integration tests.\newline\cite{evoxc2026} & \texttt{2b289f034f6f}\hfill\href{https://github.com/EMI-Group/evoxc/blob/2b289f034f6f8e16b6c9b55932dfed76cc54690e/tests/transforms/test_batch_functions_integration.py}{\sffamily GitHub source}\newline\href{https://github.com/EMI-Group/evoxc/blob/2b289f034f6f8e16b6c9b55932dfed76cc54690e/tests/transforms/test_batch_functions_integration.py}{\nolinkurl{tests/transforms/test_batch_functions_integration.py}} \\[7pt]
\hypertarget{evidence:evoxcsync2026}{E18} & XLA synchronization: completion-event lifecycle tests.\newline\cite{evoxc2026} & \texttt{2b289f034f6f}\hfill\href{https://github.com/EMI-Group/evoxc/blob/2b289f034f6f8e16b6c9b55932dfed76cc54690e/tests/backends/test_xla_synchronize.py}{\sffamily GitHub source}\newline\href{https://github.com/EMI-Group/evoxc/blob/2b289f034f6f8e16b6c9b55932dfed76cc54690e/tests/backends/test_xla_synchronize.py}{\nolinkurl{tests/backends/test_xla_synchronize.py}} \\[7pt]
\hypertarget{evidence:evoxcperf2026}{E19} & EvoX Compiler v0.1.0: benchmark records in CONTEXT.md.\newline\cite{evoxc2026} & \texttt{2b289f034f6f}\hfill\href{https://github.com/EMI-Group/evoxc/blob/2b289f034f6f8e16b6c9b55932dfed76cc54690e/CONTEXT.md}{\sffamily GitHub source}\newline\href{https://github.com/EMI-Group/evoxc/blob/2b289f034f6f8e16b6c9b55932dfed76cc54690e/CONTEXT.md}{\nolinkurl{CONTEXT.md}} \\[7pt]
\hypertarget{evidence:threeexamples2026}{E20} & three.js r184 examples.\newline\cite{threeexamples2026} & \texttt{r184}\hfill\href{https://github.com/mrdoob/three.js/tree/r184/examples}{\sffamily GitHub source}\newline\href{https://github.com/mrdoob/three.js/tree/r184/examples}{\nolinkurl{examples}} \\[7pt]
\hypertarget{evidence:mesahelper}{E21} & Benchmark timing and checksum defaults.\newline\cite{mesarepo2026} & \texttt{2ad071e658f8}\hfill\href{https://github.com/EMI-Group/genesis-demo-mesa-rs/blob/2ad071e658f8d4de21aa9f0e906592c364e10023/scripts/common.py}{\sffamily GitHub source}\newline\href{https://github.com/EMI-Group/genesis-demo-mesa-rs/blob/2ad071e658f8d4de21aa9f0e906592c364e10023/scripts/common.py}{\nolinkurl{scripts/common.py}} \\[7pt]
\end{longtable}
\endgroup
\clearpage

\section{Experimental synthesis and follow-up protocols}
\subsection{Formation, continuation and redevelopment}
The studies examine what is inherited during formation, continuation and redevelopment, and how later work is assessed. Table~\ref{tab:falsification-matrix} brings these regimes together by linking each setup to its observations and demonstrated capabilities. Sections~\ref{sec:compiler-formation} to \ref{sec:webgl} provide the underlying measurements.

\begin{table}[!htbp]\small\centering
\caption{Experimental setup, evidence and demonstrated capabilities.}\label{tab:falsification-matrix}
\begin{tabular}{@{}>{\raggedright\arraybackslash}p{0.14\linewidth}>{\raggedright\arraybackslash}p{0.22\linewidth}>{\raggedright\arraybackslash}p{0.28\linewidth}>{\raggedright\arraybackslash}p{0.28\linewidth}@{}}

\toprule
Experiment & Experimental setup & Empirical evidence in this study & Established capability scope \\
\midrule

Compiler formation & Operator specification defines language, compiler organization, target architectures and test sources & Starting repository with no compiler code, a reconstructed lineage tree of 1,015 records, 248,989 tracked lines and seven verification families & Demonstrates de novo formation at scale of a working C compiler across 1,019 transient episodes \\
Compiler continuation & Shared ancestral repository and controller limits; one run per model with independently realized resource use & Shared ancestral GLM~5.2 compiler world and identical continuation prompt, evaluated across GLM~5.2 and DeepSeek V4 Flash lineages & Demonstrates continuation across agent turnover and replacement of GLM~5.2 with DeepSeek V4 Flash \\
MESA redevelopment & Selected functionality across 13 modules; six workloads evaluated under recorded build configurations & Thirteen mapped Fortran module directories migrated into 13 Rust crates, audited across six workloads (25 runs each) and a burn proxy check of 40 runs & Demonstrates scientific redevelopment that preserves numerical behaviour under workload checksum and pointwise component checks \\
WebGL formation & Software renderer with browser-based conformance manifests & 2,017/2,019 pages passed; two WebGL2 repairs linked to commits and subtest outcomes & Demonstrates graphics formation and recovery of API ordering and image precision failures \\
\bottomrule
\end{tabular}
\end{table}

\clearpage
\subsection{Validation outcomes and archive coverage}
Alongside these capabilities, the records specify which checks passed, which cases timed out and how much of each run was archived. Table~\ref{tab:failure-cases} brings these outcomes and coverage measures together across the four studies, retaining their respective reporting conventions.

\begin{table}[H]
\centering\small
\caption{Validation outcomes, archive coverage and measurement conventions.}\label{tab:failure-cases}
\begin{tabular}{@{}>{\raggedright\arraybackslash}p{0.21\linewidth}>{\raggedright\arraybackslash}p{0.31\linewidth}>{\raggedright\arraybackslash}p{0.38\linewidth}@{}}

\toprule
Experiment & Recorded outcome & Reporting convention \\
\midrule

Compiler formation & Four direct missing parent records; four of 36 evaluated LLVM cases did not pass; 93 of 100 Csmith programs passed; seven timed out under both compilers & Missing parent records are excluded from lineage reconstruction and classified work allocation; raw archive totals retain all records \\
Compiler continuation & GLM~5.2 passed 1,445/1,448 cases on its retained LLVM SingleSource manifest & The retained set follows the final runner; comparison and collection changes are detailed in Section~\ref{sec:continuation-evaluation-changes} \\
Continuation archives & Archive coverage is 89.7\% for ancestral GLM construction, 99.0\% for GLM continuation and 94.4\% for DeepSeek continuation & Lineage topology, role distribution, depth and concurrency profiles are computed strictly from available archived records \\
MESA numerical checks & Four of six audited workloads exhibit small nonzero relative checksum discrepancies ($5.1\times10^{-15}$ to $3.1\times10^{-9}$) & Checksum discrepancies are reported per workload within the specified module coverage \\
MESA timing checks & One noisy interleaved burn run favoured Fortran ($0.82\times$), whereas the dedicated pinned batch of 40 runs favoured Rust ($1.23\times$ median) & Batch specific ratios retain their associated host conditions and build configuration \\
MESA line counts & 89,946 physical Rust lines: 67,374 library, 19,955 crate test and 2,617 root test and benchmark lines & Physical lines include comments, blank lines and unterminated final lines; the 13 mapped crates total 87,329 lines \\
WebGL conformance & Two nonpassing WebGL1 pages; 1,184/1,184 WebGL2 pages passed & Page totals retain both failures; repair-level subtest counts and visual subsets use separate denominators \\
WebGL development interval & 65.9 hours between first and last agent commits & Repository history span; model token cost of US\$36 is project-reported \\
\bottomrule
\end{tabular}
\end{table}

\Needspace{6\baselineskip}
\subsection{Planned component intervention for noncode records}
The observed trajectories motivate a planned intervention on inherited noncode records, $\operatorname{Dev}(v)$, including directory contracts, architectural invariants and provenance ledgers. To examine their contribution, the intervention will compare two accepted versions with identical executable code $\operatorname{Code}(v)$ but different noncode records under the same future objective $u$. Its outcome vector $Y$ will record task completion, regressions on the inherited test suite and resource use. The hypothesis is that changing the inherited noncode records changes these outcomes:
\begin{equation}
\begin{aligned}
\operatorname{Code}(v^{A})&=\operatorname{Code}(v^{B}), &
\operatorname{Dev}(v^{A})&\neq\operatorname{Dev}(v^{B}),\\
P\!\left(Y\mid(v^{A},p),u,B\right)&\neq P\!\left(Y\mid(v^{B},p),u,B\right),
\end{aligned}
\label{eq:supp-causal-dev}
\end{equation}
under matched repository path $p$, foundation model, toolchain, verification oracle and resource ceiling $B$.

\end{document}